\documentclass[final,3p,times,11pt,sort&compress]{elsarticle}
\usepackage{amssymb}
\usepackage{amsmath}
\usepackage{color}
\usepackage{xspace}
\usepackage{multirow}

\usepackage[utf8x]{inputenc}
\graphicspath{{./}}

\journal{Physics Letters B}

\begin{document}

\begin{frontmatter}

\title{Heavy flavor jet production and substructure in electron-nucleus collisions }

\author[a,b,c]{Hai Tao Li}
\ead{haitao.li@sdu.edu.cn}
\author[c]{Ze Long Liu}
\ead{zelongliu@itp.unibe.ch}
\author[c]{Ivan Vitev}
\ead{ivitev@lanl.gov}
\address[a]{School of Physics, Shandong University, Jinan, Shandong 250100, China}
\address[b]{HEP Division, Argonne National Laboratory, Argonne, Illinois 60439, USA}
\address[c]{Department of Physics and Astronomy, Northwestern University, Evanston, Illinois 60208, USA}
\address[d]{Theoretical Division, Los Alamos National Laboratory, Los Alamos, NM, 87545, USA}

\begin{abstract}
Deep inelastic scattering on nuclei at the Electron-Ion Collider will open
new opportunities to investigate the structure of matter. Heavy
flavor-tagged jets are complementary probes of the partonic composition and
transport coefficients of large nuclei, but introduce a new mass scale that
modifies the structure of parton showers and must be carefully accounted
for in perturbative calculations. In the framework of soft-collinear
effective theory with Glauber gluon interactions, we present the first
calculation of inclusive charm-jet and bottom-jet cross sections in
electron-nucleus collisions at next-to-leading order and compare them to
the reference electron-proton case. We also show predictions for the heavy
flavor-tagged jet momentum sharing distributions to further clarify the
correlated in-medium modification of jet substructure.
\end{abstract}   

\end{frontmatter}

\section{Introduction} 
In the past several years significant progress has been made toward defining
the physics of the future Electron-Ion Collider (EIC) and establishing
detector requirements that will enable the proposed measurements. The recently
released Yellow Report~\cite{AbdulKhalek:2021gbh} is an important first
step in summarizing the current status of deep inelastic scattering (DIS)
studies at the future facility, but is far from complete in terms of science
reach. In the years that lead to EIC operation its physics program will
continue to expand to reflect the pertinent new developments in theory
and phenomenology. In this letter we report one such development - the
first calculation of heavy flavor jet production and substructure in electron-nucleus
(e+A) collisions.

Theoretical studies of charm-quark jets (c-jets) and bottom-quark jets
(b-jets) in nucleus-nucleus (A+A) collisions~\cite{Huang:2013vaa,Li:2017wwc,Li:2018xuv,Kang:2018wrs,Wang:2019xey,Fan:2020ccq,Dai:2021mxb}
and related experimental measurements at the Large Hadron Collider~\cite{Chatrchyan:2013exa,Khachatryan:2015sva,Sirunyan:2016fcs,CMS:2019jis,Sheikh:2020sgf}
have become readily available. Heavy flavor-tagged jets will also soon
be studied at the Relativistic Heavy Ion Collider (RHIC)~\cite{Adare:2015kwa}
with the sPHENIX experiment. On one hand these results complement the physics
of inclusive jets dominated by light partons in heavy ion collisions and
provide alternative diagnostics of the transport properties of nuclear
matter. On the other hand there are unique aspects of quantum chromodynamics
(QCD) that can only be accessed with heavy flavor measurements. First and
foremost is the effect of heavy quark mass on parton showers dubbed generically
the ``dead cone'' effect~\cite{Dokshitzer:2001zm}. It was found to play
an important role in non-Abelian parton energy loss at small and moderate
heavy quark energies~\cite{Djordjevic:2003zk,Zhang:2003wk,Armesto:2003jh}
and in the full medium-induced splitting kernels~\cite{Kang:2016ofv,Sievert:2018imd,Sievert:2019cwq}
- the analogues of Altarelli-Parisi branching in nuclear matter. Second, in this
kinematic regime, the heavy quark mass can produce quantitatively and even
qualitatively different modification of jet observables in reactions with
nuclei in comparison to the massless case~\cite{Li:2017wwc}. For an overview
or heavy flavor physics, albeit with more emphasis on hadron production,
see e.g. Refs.~\cite{Frawley:2008kk,Andronic:2015wma,Dong:2019byy}.

At the same time, theoretical studies of heavy flavor jets at the EIC have
been extremely limited. In electron-proton (e+p) collisions c-jets produced
in charge current reactions have been proposed~\cite{Arratia:2020azl} as
a probe of the strangeness content of the proton at intermediate and large
values of Bjorken-x. The transverse spin asymmetry of back-to-back heavy
flavor-tagged jet production has been shown to be sensitive to the gluon
Sivers function~\cite{Kang:2020xgk}. Experimental feasibility studies for
heavy flavor jet measurements at the EIC have also been performed~\cite{Arratia:2020azl,Li:2021xig}.
In electron-nucleus collisions this physics is not yet developed. Ref.~\cite{Chudakov:2016otl}
explored the possibility of using the total charm production cross section
to constrain the gluon nuclear parton distribution function (nPDF). Semi-inclusive
D-meson and B-meson production in DIS on nuclei can shed light on the physics
of hadronization and differentiate between competing paradigms of hadron
attenuation in cold nuclear matter~\cite{Li:2020zbk,Li:2020sru}. In this
letter we take the studies of heavy flavor in cold nuclear matter to the
next level and present predictions for c-jet and b-jet cross section and
substructure modification in electron-nucleus collisions.

To address this problem we can take guidance from the recent calculation
of light jet rates and the jet charge in DIS on nuclei~\cite{Li:2020rqj}.
Even in the absence of nuclear matter, the heavy quark mass $m$ must be
accounted for in perturbative calculations. It was introduced in Refs.~\cite{Rothstein:2003wh,Leibovich:2003jd}
in the framework of soft-collinear effective theory (SCET)~\cite{Bauer:2001yt,Bauer:2002nz,Beneke:2002ph}
with focus on the $m/p =\lambda \ll 1 $ regime, yielding a new formulation
with finite mass corrections SCET$_{\mathrm{M}}$. The role of the heavy quark
mass in heavy flavor-tagged jet cross sections evaluated with the help
of semi-inclusive jet functions (SiJFs) was also understood~\cite{Dai:2018ywt,Li:2018xuv}.
In addition to the logarithms of the ratio of the hard scale $\mu $ to
the jet scale $p_{T} R$ that should be resumed for small radii $R$, logarithms
of the ratio of the jet scale to the heavy quark mass have also been accounted
for. For reactions with nuclei, SCET$_{\mathrm{M}}$ has been generalized to
describe parton shower interactions in matter via the exchange of off-shell
Glauber gluons~\cite{Kang:2016ofv}. This development, which resulted in
the derivation of the in-medium splitting functions for heavy quarks, has
been complemented by the re-analysis of parton branching using the formalism
of lightcone wavefunctions~\cite{Sievert:2018imd}, confirming earlier results
and allowing to compute splittings in QCD media to any order in the opacity
of matter. Last but not least, the contribution of medium-induced parton
showers to the SiJFs for inclusive jets~\cite{Kang:2017frl} and charm-quark
jets / bottom-quark jets~\cite{Li:2018xuv} has been derived, the latter
being particularly relevant to this work. Here we show how this approach
can be applied to deep inelastic scattering, and in particular to e+A reactions.
We will complement the calculation of heavy flavor-tagged jet quenching
in cold nuclear matter with the evaluation of the c-jet and b-jet soft-dropped
momentum sharing distributions~\cite{Ilten:2017rbd}. This observable is
especially illuminating since its modification in QCD matter can be quite
different at small and moderate transverse momenta relative to large ones,
depending on the quark mass~\cite{Li:2017wwc}.

The rest of this letter is organized as follows: in Section~\ref{formalism}
we present the theoretical formalism for calculating the b-jet and c-jet
cross sections and the soft-dropped momentum sharing distributions in e+p
and e+A reactions. Phenomenological results for DIS on a gold (Au) nucleus
are shown in Section~\ref{results}. We conclude in Section~\ref{conclusions}.

\section{Theoretical Framework}\label{formalism}
\subsection{Semi-inclusive jet cross sections at the EIC}

The factorization formula for the cross section for semi-inclusive jet
production in collinear leading-twist perturbative QCD can be written as~\cite{Hinderer:2015hra}:
%
%e1 #&#
\begin{equation}
\label{eq:NLOform}
\begin{aligned}
E_{J} &\frac{d^{3} \sigma }{d^{3} P_{J}} =\frac{1}{S} \sum _{i, f}
\int _{0}^{1} \frac{d x}{x} \int _{0}^{1} \frac{d z}{z^{2}} f^{i / N}(x,
\mu ) \, J_{J_{Q} / f}(z, p_{T} R, m, \mu )\, \Big [\hat{\sigma }_{i
\rightarrow f} +f_{\mathrm{ren}}^{\gamma /\ell }\left (\frac{-t}{s+u},\mu
\right )\hat{\sigma }^{\gamma i \to f}\Big ] \, .
\end{aligned}
\end{equation}
Here $ f^{i / N}$ is the parton distribution function (PDF) of parton
$i$ in nucleon $N$ and $J_{J_{Q} / f}$ is the SiJF from parton $f$ to jet
$J_{Q}$ containing heavy flavor. $z$ and $m$ denote the momentum fraction
taken by the jet and the mass of the heavy flavor parton, respectively.
$p_{T}$ and $R$ are the transverse momentum and radius of the semi-inclusive
jet. $\hat{\sigma }^{i\to f}$ denotes the cross section for lepton-parton
scattering with initial-state parton $i$ and final-state parton $f$. Lastly
$s$, $t$, $u$ are the partonic Mandelstam variables defined as
$s=(k+l)^{2}$, $t=(k-p)^{2}$ and $u=(l-p)^{2}$, where $l^{\mu }$,
$k^{\mu }$ and $p^{\mu }$ are the momenta of the incoming lepton, the incoming
parton and the fragmenting parton, respectively. Because kinematic constraints
on the scattered lepton are not employed in jet production at the EIC,
events with forward lepton scattering can be selected. In this case, the
hard process can be described by an incoming quasi-real photon scattering:
$\gamma q\to q(g)$, $\gamma q\to g(q)$, $\gamma g\to q({\bar{q}})$, which
contributes to the cross section starting at order
$\alpha _{\mathrm{EM}}^{2} \alpha _{s}$. Quasi-real photons originate from the
incoming lepton and can be accurately described by the well known Weizs\"{a}cker-Williams
(WW) distribution with a perturbative distribution function
$f_{\mathrm{ren}}^{\gamma /\ell }\left (y,\mu \right )$~\cite{vonWeizsacker:1934nji,Williams:1934ad,Bawa:1989bf,Frixione:1993yw}.
The analytical expressions for $\hat{\sigma }^{i\to f}$,
$\hat{\sigma }^{\gamma i\to f}$ and
$f_{\mathrm{ren}}^{\gamma /\ell }\left (y,\mu \right )$ up to
${\mathcal{O}}(\alpha _{\mathrm{EM}}^{2} \alpha _{s})$ can be found in~\cite{Hinderer:2015hra}.
We note that there is also a resolved photon contribution, related to the
partonic content of the $\gamma $~\cite{Uebler:2017glm}. Formally, it starts
at ${\mathcal{O}}(\alpha _{\mathrm{EM}}^{2} \alpha _{s}^{2})$ which is homogeneous
with NNLO QCD corrections~\cite{Boughezal:2018azh} and contributes at relatively
small transverse momenta. Furthermore, the overall cross section normalization
of high transverse momentum jets will cancel in the nuclear modification
ratio that we study in Section~\ref{results}. For these reasons, we did
not consider the resolved photon component here and defer its investigation
to future studies.

The NLO SiJFs for heavy flavor jets can be found in Ref.~\cite{Dai:2018ywt}.
The evolution of heavy flavor SiJFs is briefly reviewed below. Since the heavy
quark mass does not affect the ultraviolet (UV) behavior of diagrams, the
evolution of heavy-flavor SiJFs obeys DGLAP-like equations similar to the
ones for light-flavor SiJFs. The renormalization-group equation (RGE) is
given by
%
%e2 #&#
\begin{align}
\label{eq:dglap}
&\frac{d}{d\ln \mu ^{2}} \left (
\begin{array}{c}
J_{J_{Q}/s} (x,\mu )
\\
J_{J_{Q}/g} (x,\mu )
\end{array} \right ) = \frac{\alpha _{s}}{2\pi } \int _{x}^{1}
\frac{dz}{z} \left (
\begin{array}{c@{\quad }c}
P_{qq}(z) & 2 P_{gq(z)}
\\
P_{qg}(z) & P_{gg}(z)
\end{array} \right )\cdot \left (
\begin{array}{c}
J_{J_{Q}/s} (x/z,\mu )
\\
J_{J_{Q}/g} (x/z,\mu )
\end{array} \right ),
\end{align}
where $s=Q+\bar{Q}$. Here, $P_{ij}$ are usual Altarelli-Parisi splitting
functions. To solve the above RGE, we work in Mellin moment space following
the method outlined in \cite{Vogt:2004ns}
%
%e3 #&#
\begin{equation}
J_{Q/g}(N)=\int _{0}^{1} dz\, z^{N-1} J_{Q/g}(z) \, .
\end{equation}
The solution for the jet function in this space is given by
%
%e4 #&#
\begin{equation}
\label{eq:DGLAP3}
\begin{aligned}
\begin{pmatrix}
J_{J_{Q}/s} (N,\mu )
\\
J_{J_{Q}/g} (N,\mu )
\end{pmatrix}
=& \left [ e_{+}(N) \left (
\frac{\alpha _{s}(\mu )}{\alpha _{s}(\mu _{J})} \right )^{-r_{-}(N)} +
e_{-}(N) \left (\frac{\alpha _{s}(\mu )}{\alpha _{s}(\mu _{J})}
\right )^{-r_{+}(N)} \right ] \cdot
\begin{pmatrix}
J_{J_{Q}/s} (N,\mu _{J})
\\
J_{J_{Q}/g} (N,\mu _{J})
\end{pmatrix}
,
\end{aligned}
\end{equation}
where $r_{+}(N)$ and $r_{-}(N)$ are the larger and smaller eigenvalue of
the leading-order singlet evolution matrix,
%
%e5 #&#
\begin{equation}
\begin{aligned}
r_{\pm }(N)=&\frac{1}{2\beta _{0}}\Bigg [P_{qq}(N)+P_{gg}(N) \pm
\sqrt{\left (P_{qq}(N)-P_{gg}(N)\right )^{2} +4 P_{qg}(N) P_{gq}(N)}
\Bigg ] \, .
\end{aligned}
\end{equation}
The projector matrices $e_{\pm }(N)$ in~(\ref{eq:DGLAP3}) are defined as
%
%e6 #&#
\begin{equation}
\begin{aligned}
e_{\pm }(N)=&\frac{1}{r_{\pm }(N) - r_{\mp }(N)}
\begin{pmatrix}
P_{qq}(N)-r_{\mp }(N) & ~2 N_{f} P_{gq}(N)
\\
P_{qg}(N) & ~P_{gg}(N)-r_{\mp }(N)
\end{pmatrix}
\, .
\end{aligned}
\end{equation}
Eventually, the evolved SiJFs in $z$-space can be obtained by performing
an inverse Mellin transformation.
%
%e7 #&#
\begin{equation}
J_{J_{Q}/g}(z, \mu )=\frac{1}{2\pi i}\int _{{\mathcal{C}}_{N}} dN\, z^{-N} J_{J_{Q}/g}(N,
\mu )\,,
\end{equation}
where we chose the contour in the complex $N$ plane to the right of all
the poles of $J_{J_{Q}/g}(N, \mu )$. The NLO medium corrections to the
$Q\to J_{Q}$ and $g\to J_{Q}$ SiJFs are similar to the case of heavy-ion
collisions~\cite{Li:2018xuv} and can be written as
%
%e8 #&#
\begin{align}
J^{\text{med},(1)}_{J_{Q}/Q}(z, p_{T} R,m, \mu )
\nonumber
=& \int _{z(1-z) p_{T} R }^{\mu } d^{2} q_{\perp }f^{\mathrm{med}}_{Q\to Q+g}(z,
m, q_{\perp }) - \delta (1-z) \int _{0}^{1} dx \int _{x(1-x) p_{T} R}^{
\mu } d^{2} q_{\perp }f^{\mathrm{med}}_{Q\to Q+g}(x, m, q_{\perp })
\nonumber
\\
= & \left [\ \int _{z(1-z) p_{T} R}^{\mu } d^{2} q_{\perp }f^{\mathrm{med}}_{Q
\to Q+g}(z, m, q_{\perp }) \right ]_{+}~,
\end{align}
and
%
%e9 #&#
\begin{align}
J^{\text{med},(1)}_{J_{Q}/g}(z, p_{T} R,m, \mu ) = & \left [\ \int _{z(1-z)
p_{T} R}^{\mu } d^{2} q_{\perp }f^{\mathrm{med}}_{g\to Q+\bar{Q}}(z, m,
q_{\perp }) \right ]_{+}~ + \int _{z(1-z) p_{T} R}^{\mu }
d^{2} q_{\perp }f^{\mathrm{med}}_{g\to Q+\bar{Q}}(z, m, q_{\perp }) ~,
\end{align}
respectively. Here $f^{\mathrm{med}}_{i\to j+k}$ is the medium induced splitting
kernel. The scale $\mu $ was introduced as a UV cut-off for the medium
corrections~\cite{Kang:2017frl,Li:2018xuv} which is set to be the jet transverse
momentum in this work. In the presence of a QCD medium, it was demonstrated
that the vacuum splitting must be replaced by the full splitting kernels
for each possible branching channel
%
%e10 #&#
\begin{align}
\frac{dN^{\mathrm{full}}}{dz d^{2} q_{\perp }} =
\frac{dN^{\mathrm{vac}}}{dz d^{2} q_{\perp }} + f^{\mathrm{med}}(z, m, q_{\perp })
\; .
\label{fullsplit}
\end{align}
Hence, the full in-medium SiJFs are obtained as
%
%e11 #&#
\begin{align}
J_{J_{Q}/i} = J_{J_{Q}/i}^{\mathrm{vac}} + J_{J_{Q}/i}^{\mathrm{med}} \, ,
\end{align}
where the vacuum contributions are calculated at the LL accuracy, while
only the fixed-order medium corrections are included consistently.

\subsection{Soft-dropped jet momentum sharing distribution}
Jet substructure is a promising to study the mass effects in parton shower evolution. In this work, we focus on the jet momentum sharing variable based on the “soft drop grooming”~\cite{Larkoski:2014wba}  in  1$\to $2 QCD splitting processes. It is defined as the distribution of
\begin{align}
z_g=\frac{\min(p_{T1}, p_{T2})}{p_{T1}+p_{T2}}~, \quad  z_g>z_{\rm cut} \,,
\label{zg}
\end{align}
where $p_{T1}$ and $p_{T2}$ are the transverse momenta of the subjets in a reconstructed jet. For the heavy-flavor jet, we are interested in the kinematic region where the jet energy is much larger than the heavy quark mass,  $0<m \ll p_T$.

Consider an off-shell parton of momentum
$[p^{+} ,p^{-},{\mathbf{0}}_{\perp }]$, with $p^{+}$ the large lightcone component,
that splits into two daughter partons
$[zp^{+} ,q^{2}_{\perp }/zp^{+} ,{\mathbf{q}}_{\perp }]$ and
$[(1-z)p^{+} , q^{2}_{\perp }/(1-z)p^{+},- {\mathbf{q}}_{\perp }]$. The massive
splitting kernel $Q\to Qg$, for example, in the vacuum reads
\begin{equation} \label{eq:Msp1}
\begin{aligned}
\left(\frac{dN^{\rm vac}}{dz d^2 q_\perp}\right)_{Q\to Qg} =&  \frac{\alpha_s}{2 \pi^2} \frac{C_F}{q_\perp^2+z^2m^2}   \left( 
\frac{1+(1-z)^2}{z}-\frac{2z(1-z)m^2}{q_\perp^2+z^2m^2}
\right)~. 
\end{aligned}
\end{equation}
Here $C_F$ is the quadratic Casimir invariant of the fundamental representation of SU(3).  For $q_\perp \gg m$ Eq.~(\ref{eq:Msp1}) reduces to the massless splitting functions, however when $q_\perp \leq z m$ the heavy quark mass will significantly modify the momentum sharing observable - an effect that is amplified in nuclear matter~\cite{Li:2017wwc}. If  $r_g$ is the angular separation between the subjets and $p_T$ is the transverse momentum,  then $q_\perp = z(1-z) r_g p_T$ when $r_g$ is not too large. The interesting regime discussed above can easily be reached with the moderate transverse momenta available at the EIC, especially for b-jets. Here, we start with the vacuum case to set up the stage for the jet splitting function calculation in heavy ion collisions.  After soft-drop grooming in the parton branching $i\to jk$, the $\theta_g$ and $z_g$ distribution for parton $i$ is 
\begin{align}
\left(\frac{dN^{\rm vac}}{dz_g d\theta_g}\right)_{i}  = \frac{\alpha_s}{\pi} \frac{1}{\theta_g} \sum_{j} P_{i\to j k}^{\rm vac}(z_g) \, ,
\end{align}
with $r_g=\theta_g R$, $R$ being the jet radius.
Resummation is necessary in the kinematic region with large splitting probability. It was performed to modified leading-logarithmic (MLL) accuracy in Ref.~\cite{Larkoski:2014wba}. The resummed distribution for a $i$-type jet, initiated by a 
quark or a gluon,  is 
\begin{align} \label{eq:mll}
& \frac{dN_i^{\rm vac,MLL}}{ dz_g d\theta_g} = \sum_{j} \left(\frac{dN^{\rm vac}}{dz_g d\theta_g}\right)_{i\to j k}
\exp \left[-\int_{\theta_g}^1 d\theta \int_{z_{\rm cut}}^{1/2} dz  \sum_{i} \left(\frac{dN^{\rm vac}}{dz d\theta}\right)_{i\to j k } \right]~.
\end{align}
The normalized joint probability distribution then reads
\begin{align}
p(\theta_g,z_g)\big|_{i}={\frac{dN_i^{\rm vac,MLL}}{ dz_g d\theta_g} } \; \Bigg / \; {\int_{0}^1 d\theta \int_{z_{\rm cut}}^{1/2} dz  \frac{dN_i^{\rm vac,MLL}}{ dz d\theta} }~,\, 
\label{eq:jointprob}
\end{align}
and in for e+A collisions the corresponding medium contribution must be included according to Eq.~(\ref{fullsplit}).

\section{Numerical results}\label{results}
\subsection{Semi-inclusive heavy flavor jet  cross section modification in e+A relative to e+p at the EIC}

 \begin{figure}
 	\centering
 	\includegraphics[width=0.39\textwidth]{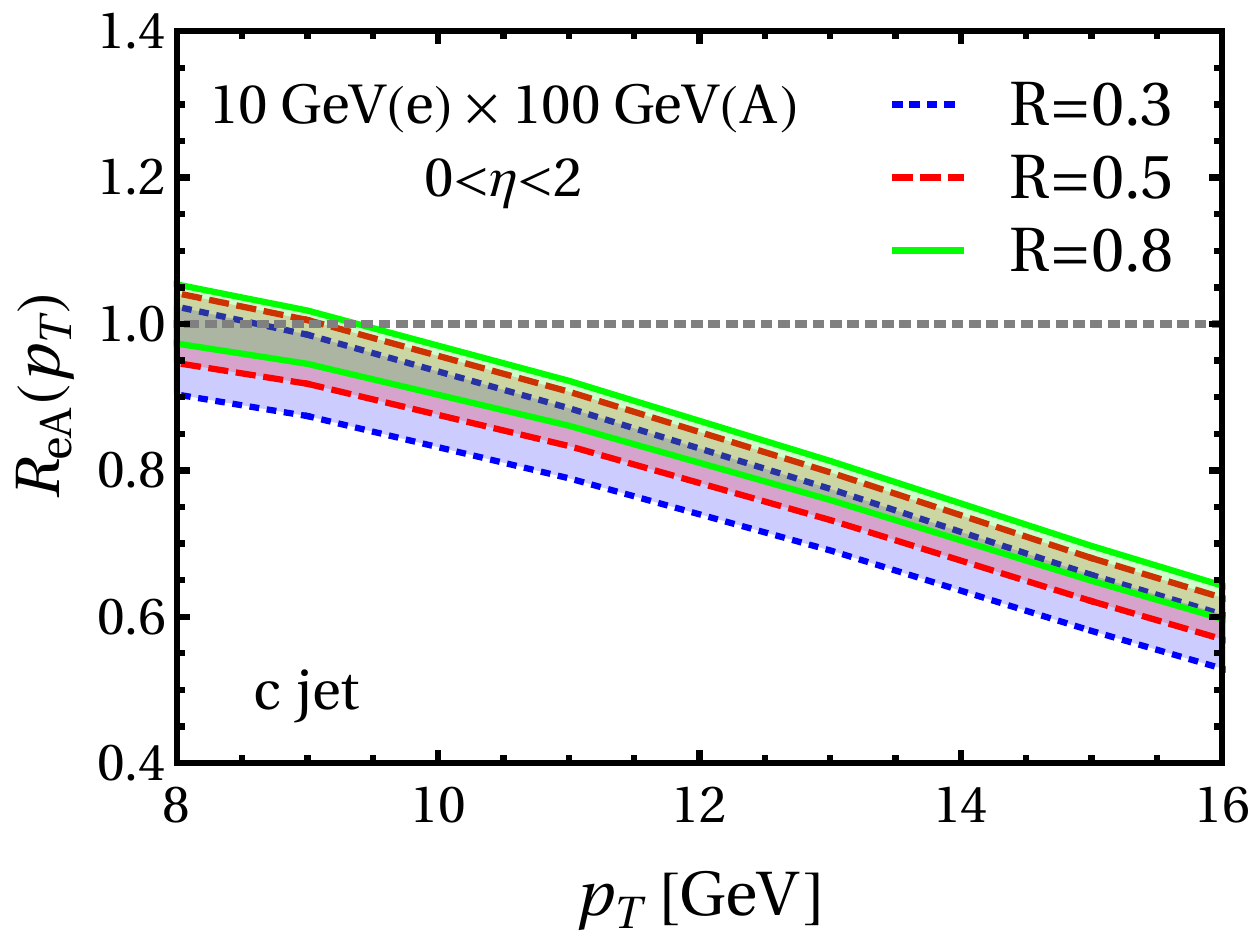}\,\,\,
 	\includegraphics[width=0.39\textwidth]{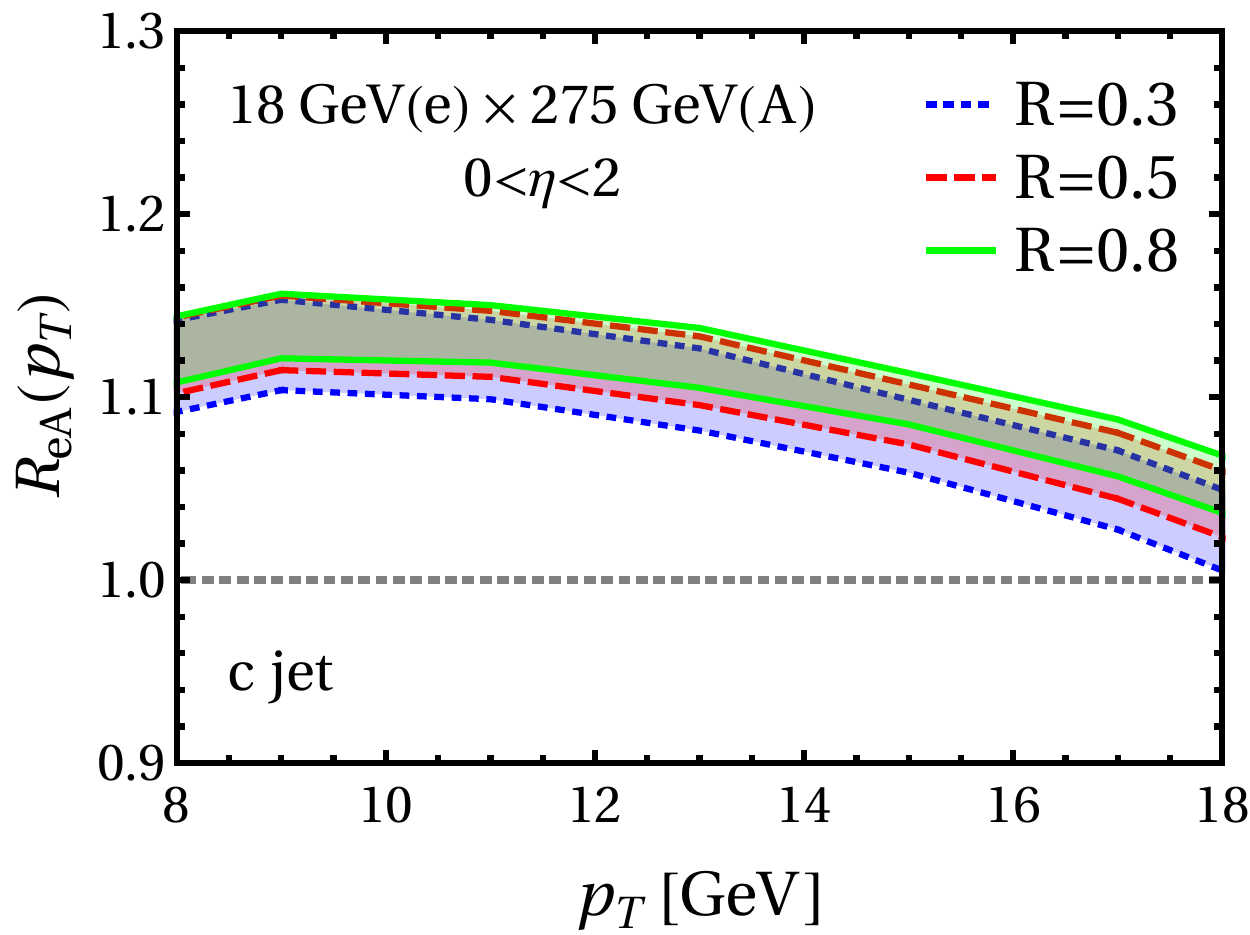} \\	
 	\includegraphics[width=0.39\textwidth]{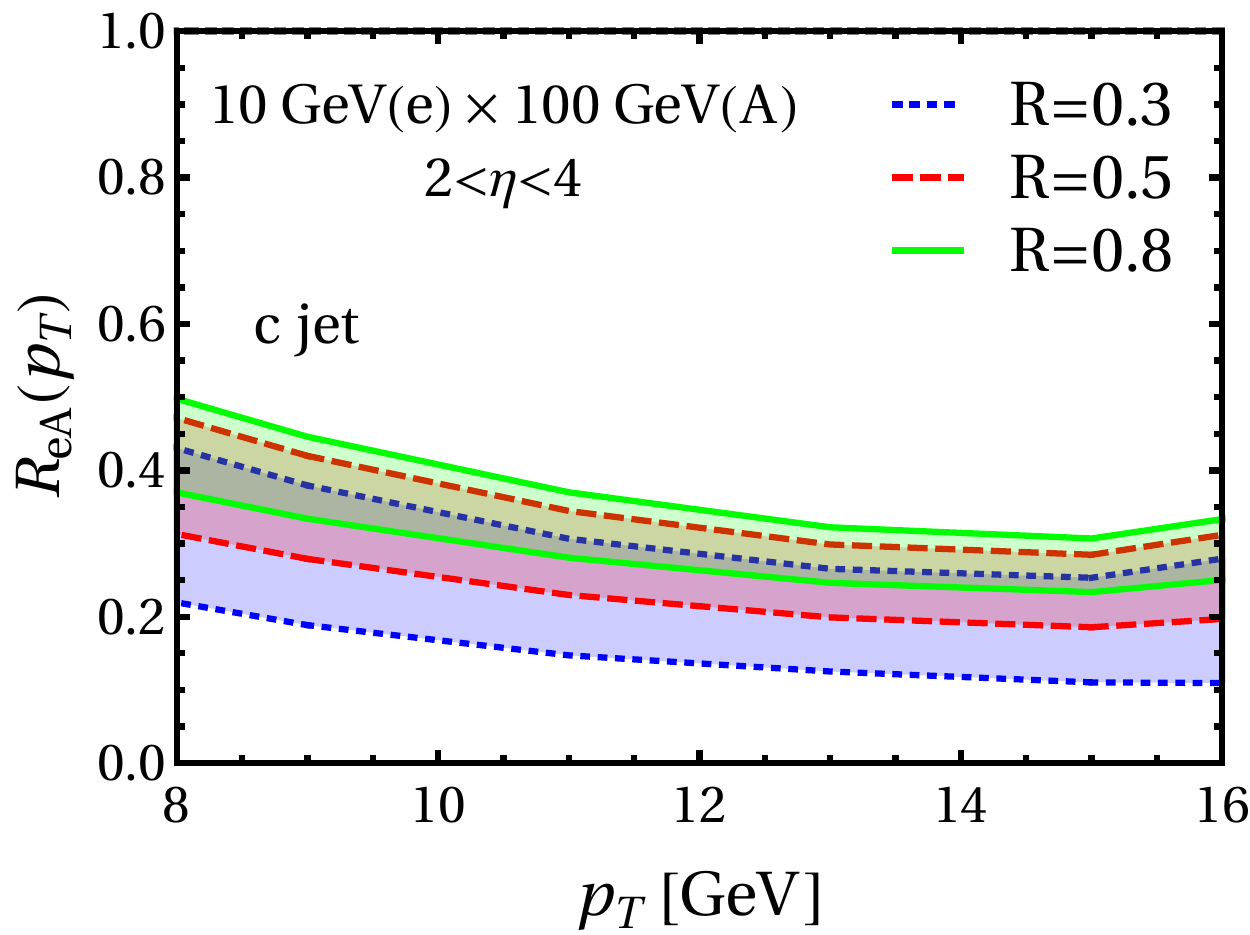}\,\,\,
 	\includegraphics[width=0.39\textwidth]{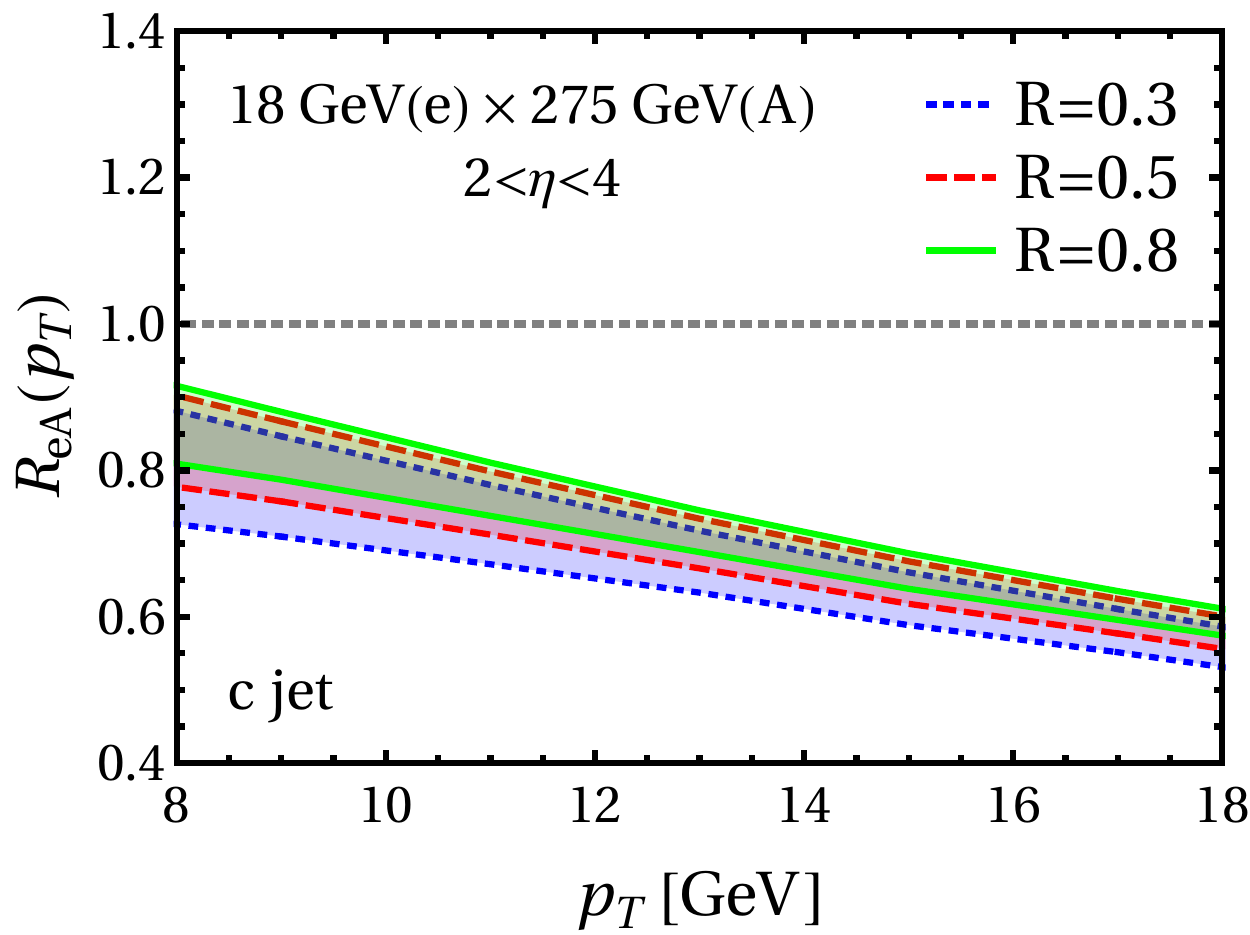} 
 	\includegraphics[width=0.39\textwidth]{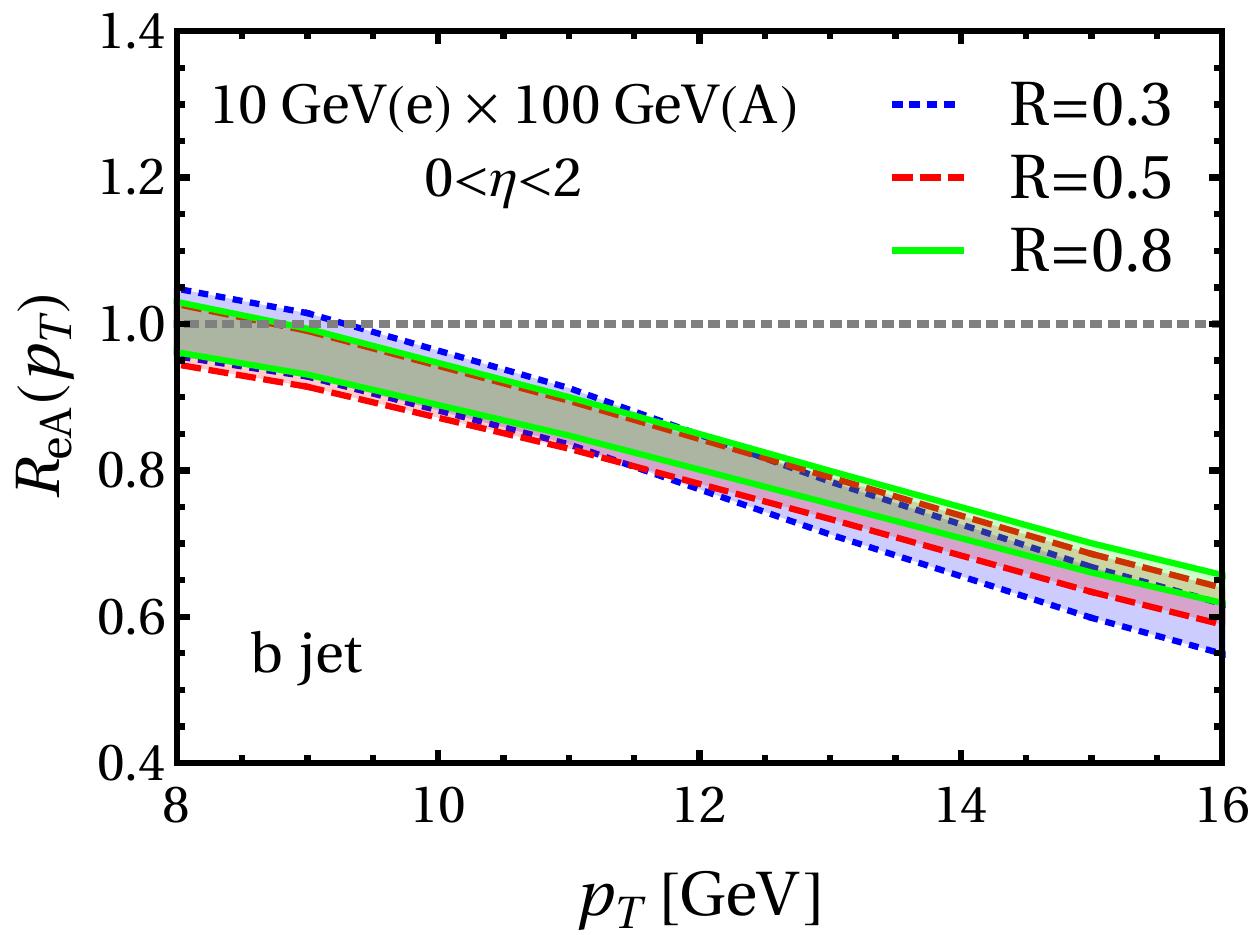}\,\,\,
 	\includegraphics[width=0.39\textwidth]{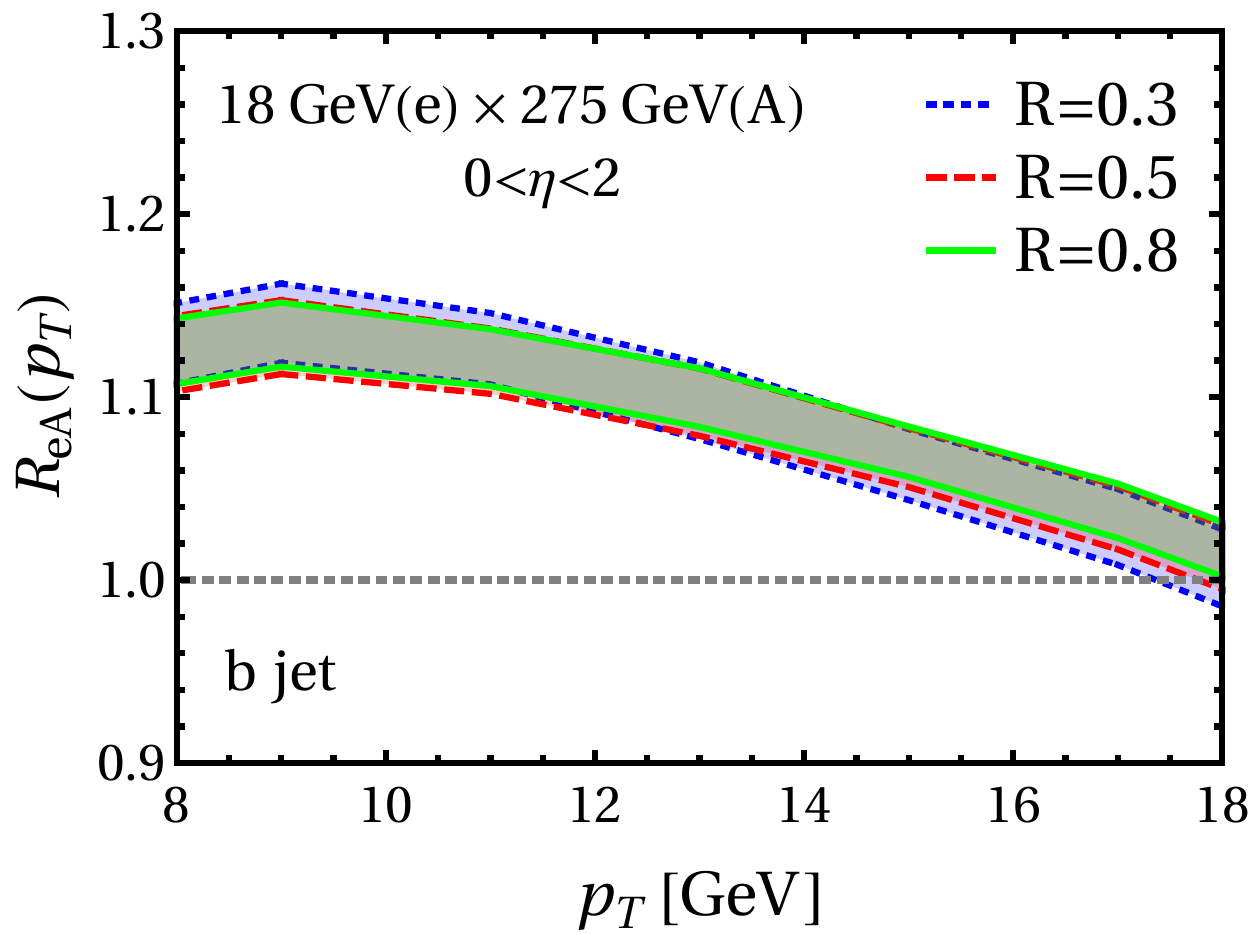} \\	
 	\includegraphics[width=0.39\textwidth]{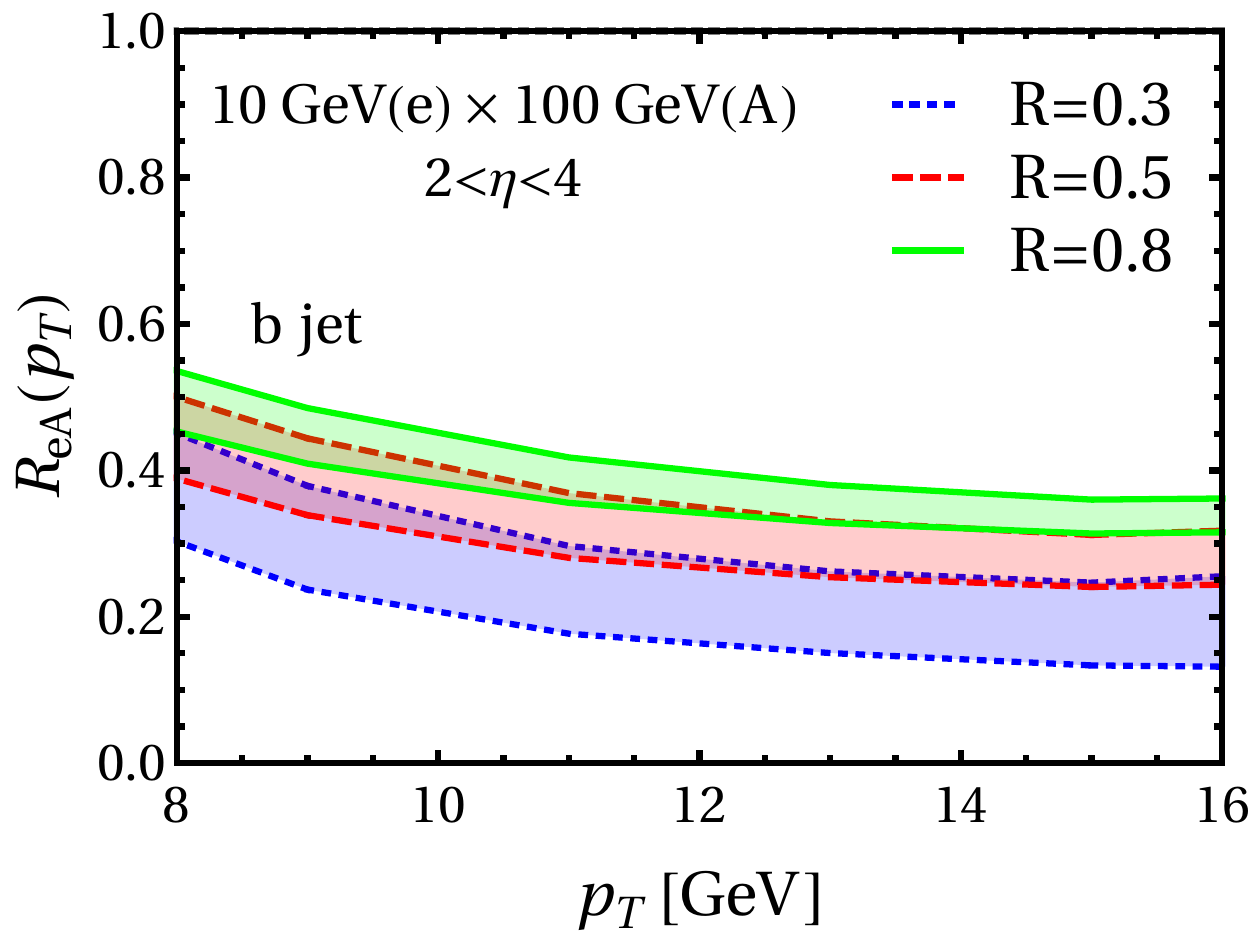}\,\,\,
 	\includegraphics[width=0.39\textwidth]{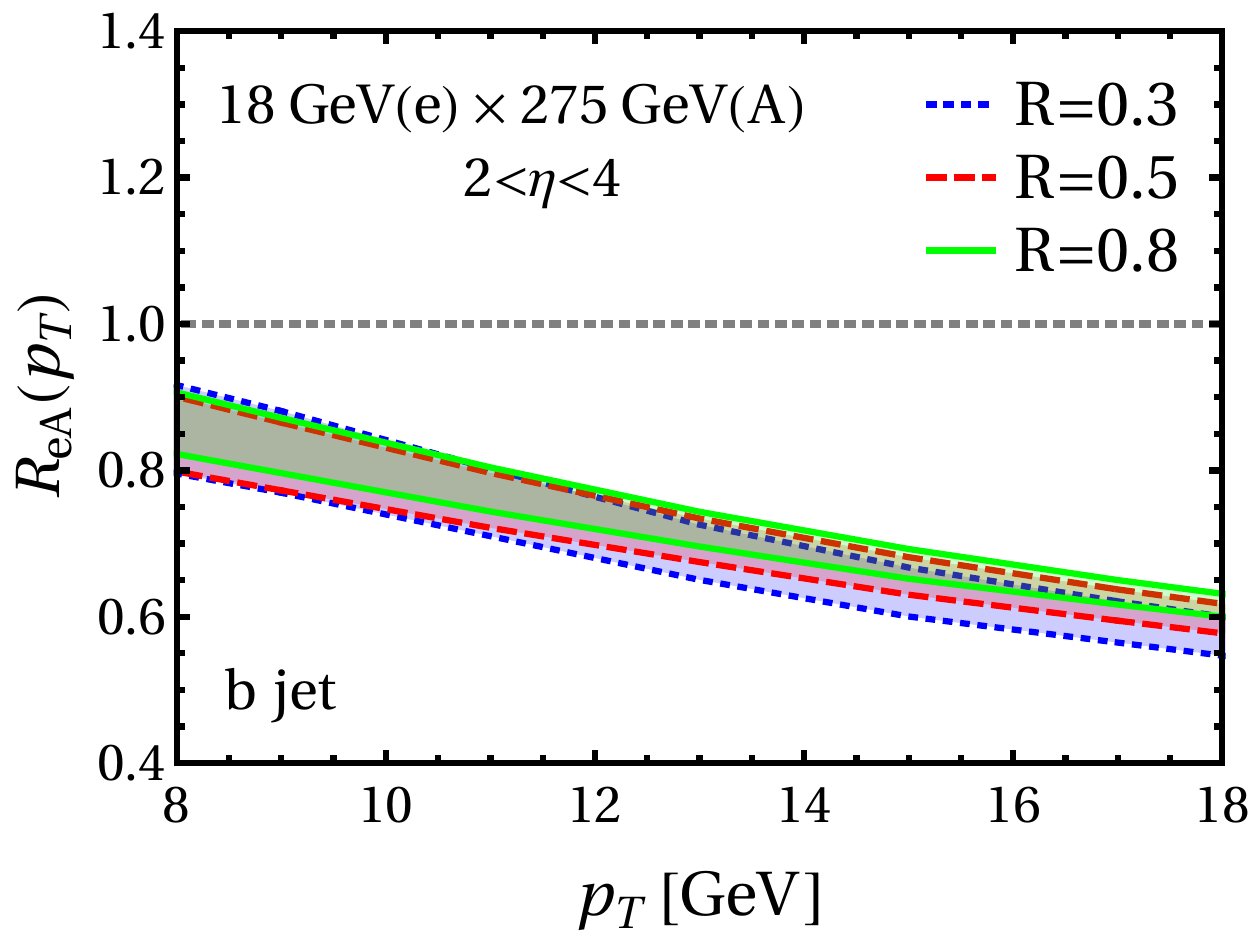} 
 	\caption{ In-medium corrections for charm-quark jets and bottom-quark jets  as a function of $p_T$ at the EIC  in two rapidity regions. Green bands (solid lines), red bands (dashed lines), and  blue bands (dotted lines) correspond to  $R=0.8$, $R=0.5$ and $R=0.3$, respectively.
 		Results for 10~GeV(e) $\times$ 100 GeV(A)  collisions are shown on the left and results for 18 GeV(e) $\times$ 275 GeV(A) collisions  are shown on the right. }
 	\label{fig:bcj_EIC}
 \end{figure}
 
In this section  we present the main result of this work  - heavy flavor jet cross section modification at the EIC.
Here we consider two benchmarks energy combinations  for electron-proton collisions (for electron-nucleus collisions, the beam energy is per nucleon): 10 GeV  (e) $\times$ 100 GeV (A) and 18 GeV (e) $\times$ 275 GeV (A). 

\begin{figure}
	\centering
	\includegraphics[width=0.39\textwidth]{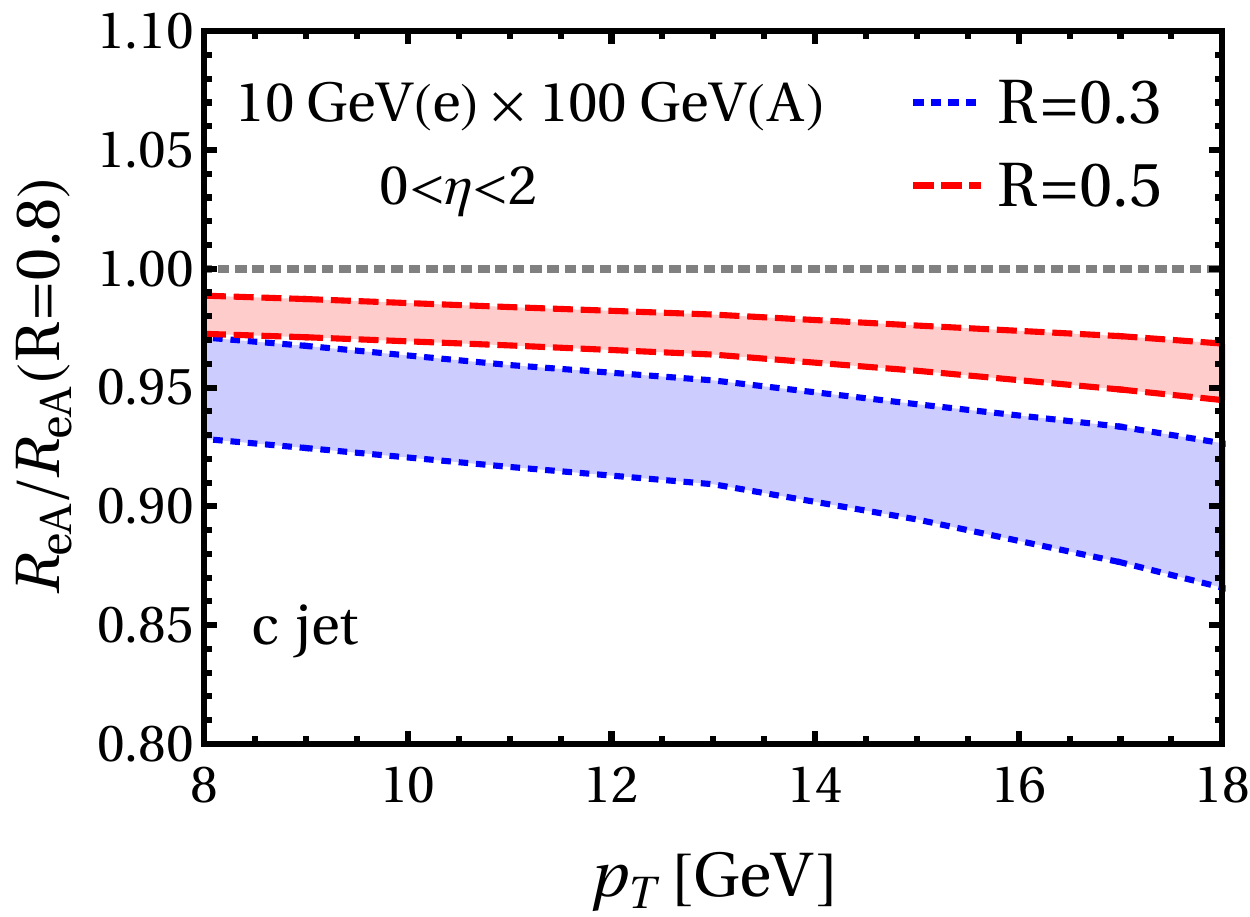}\,\,\,
	\includegraphics[width=0.39\textwidth]{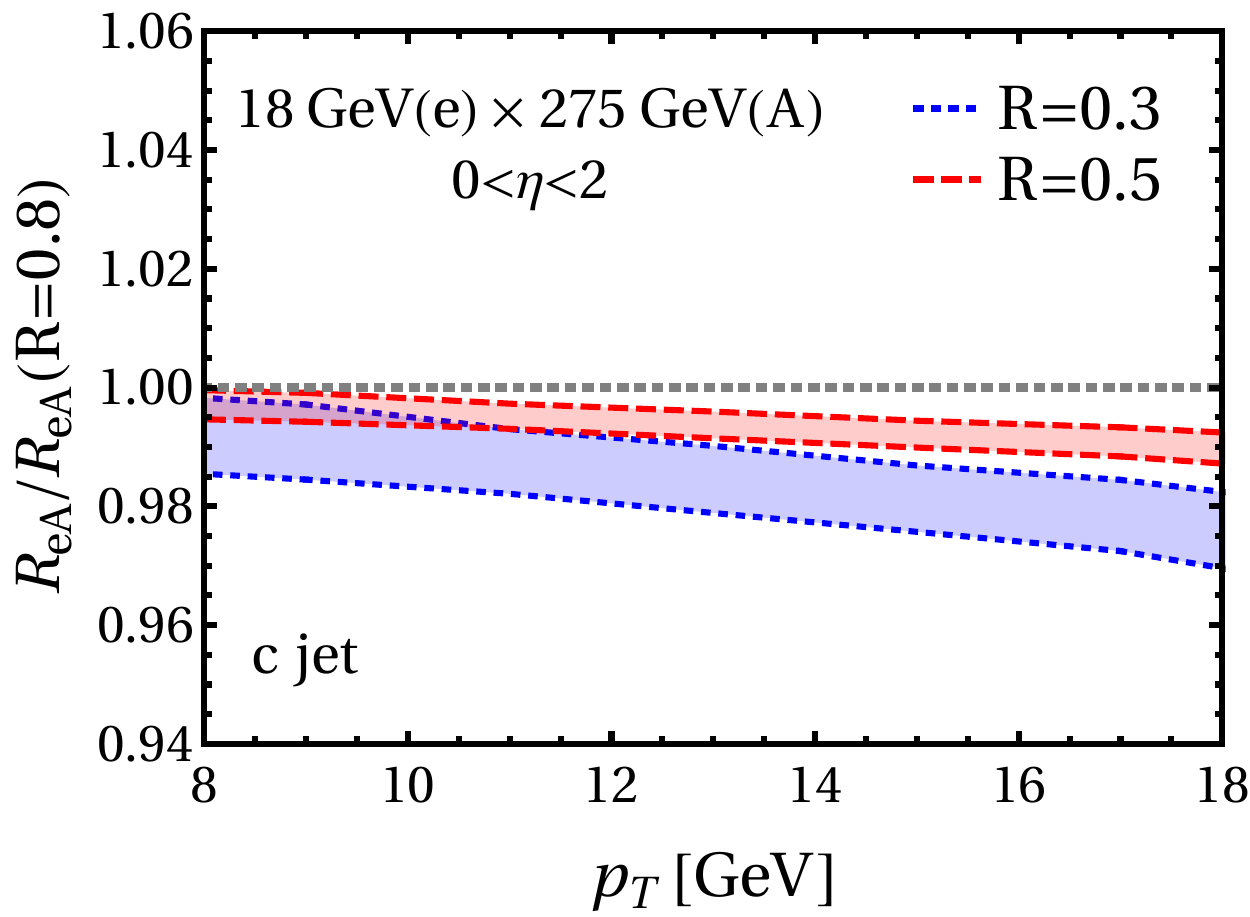} \\	
	\includegraphics[width=0.39\textwidth]{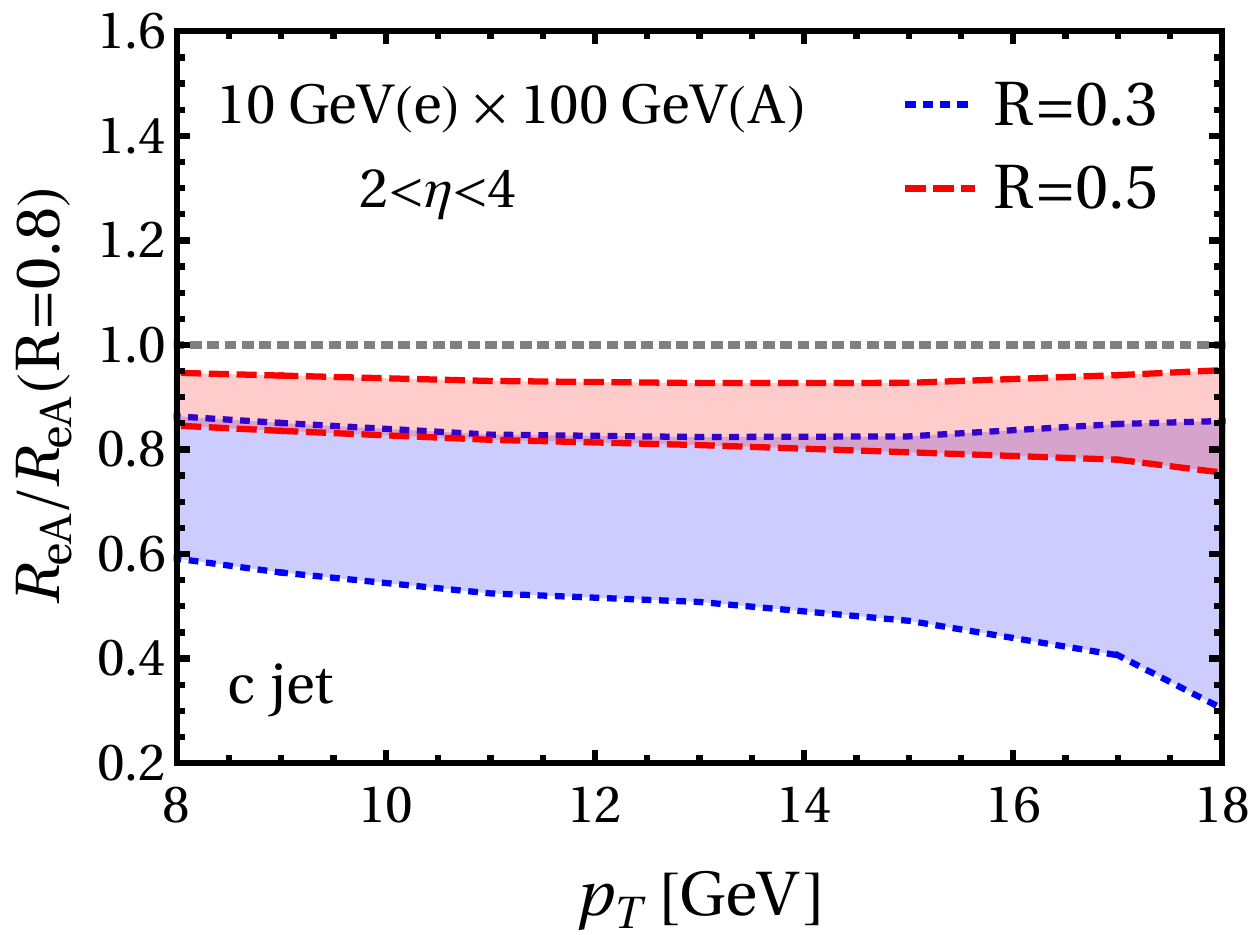}\,\,\,
	\includegraphics[width=0.39\textwidth]{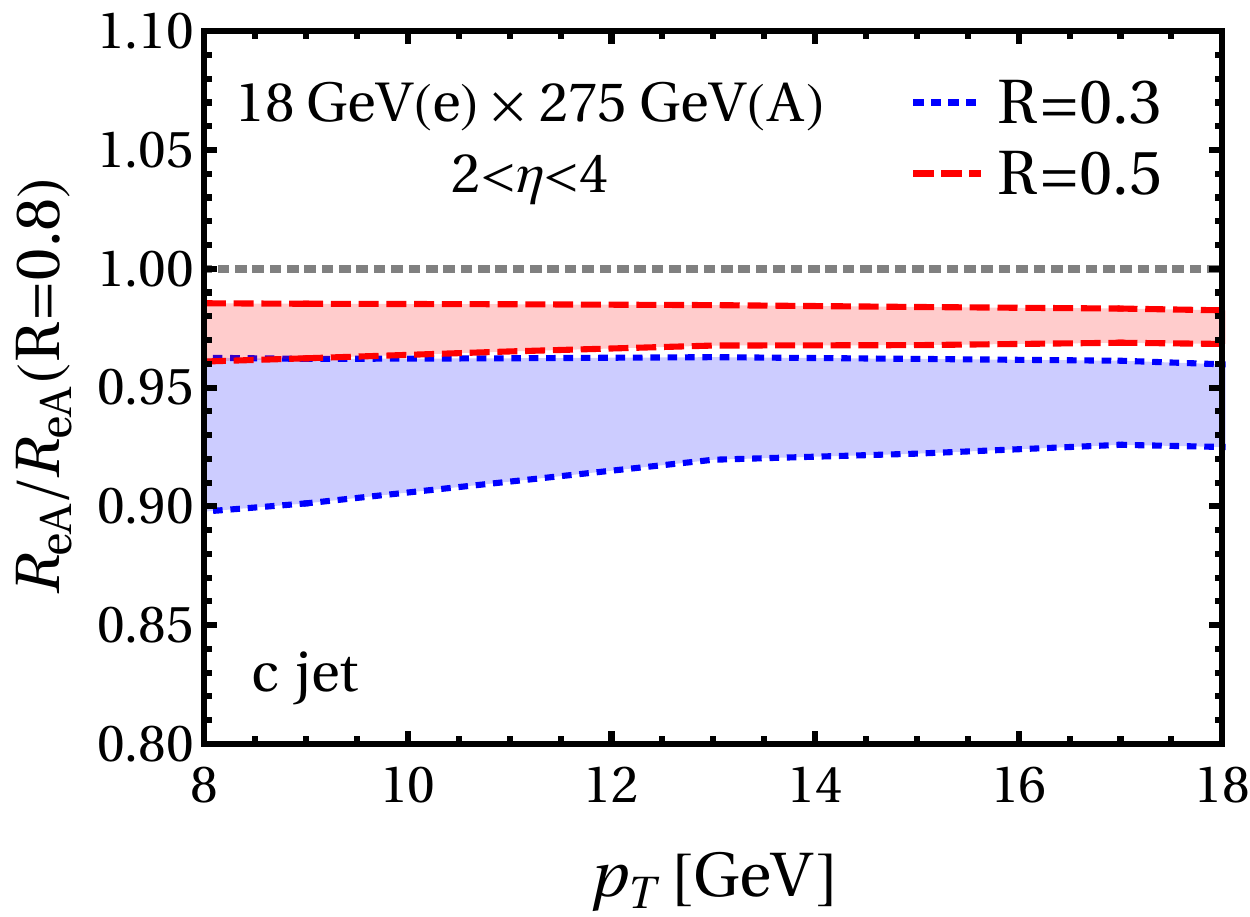} 
	\includegraphics[width=0.39\textwidth]{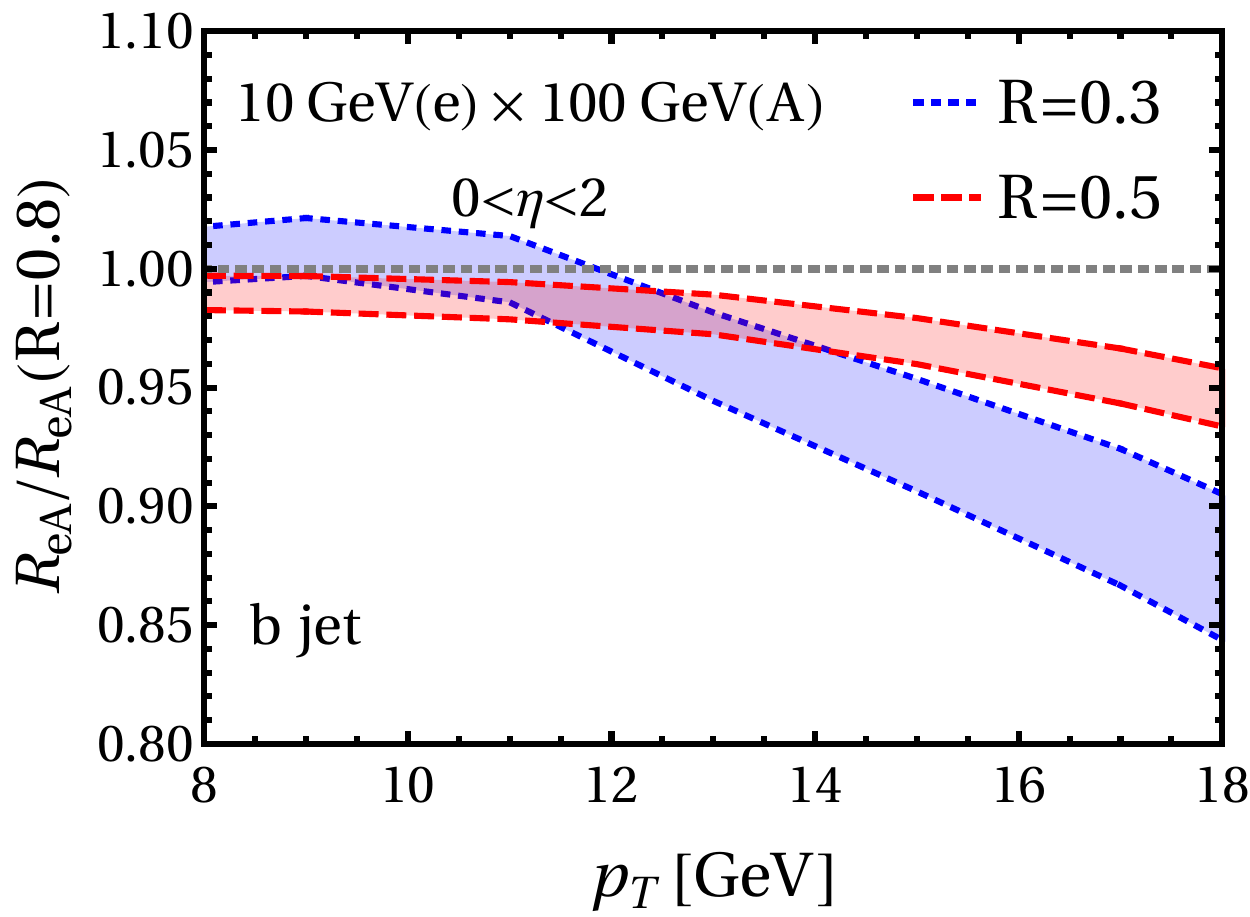}\,\,\,
	\includegraphics[width=0.39\textwidth]{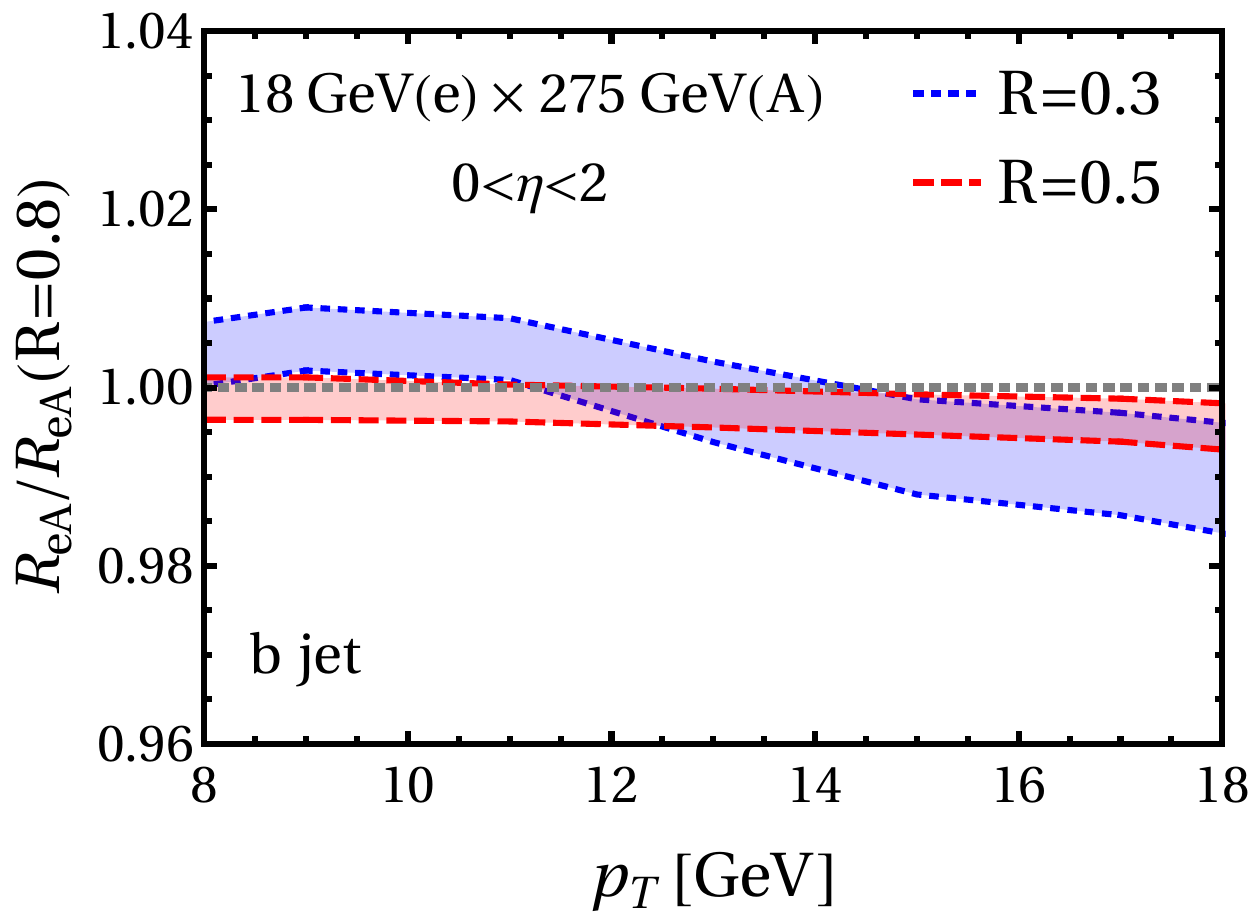} \\	
	\includegraphics[width=0.39\textwidth]{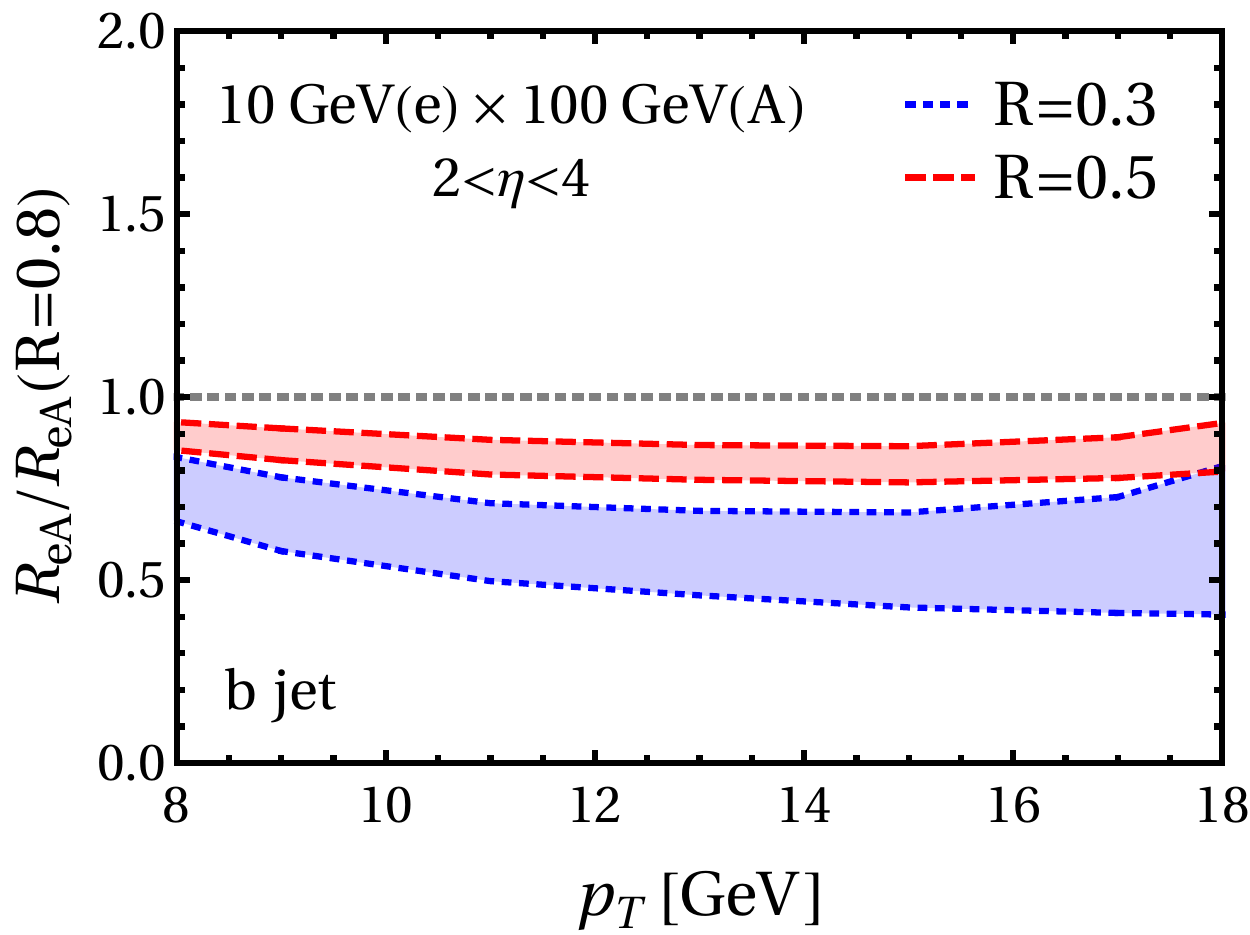}\,\,\,
	\includegraphics[width=0.39\textwidth]{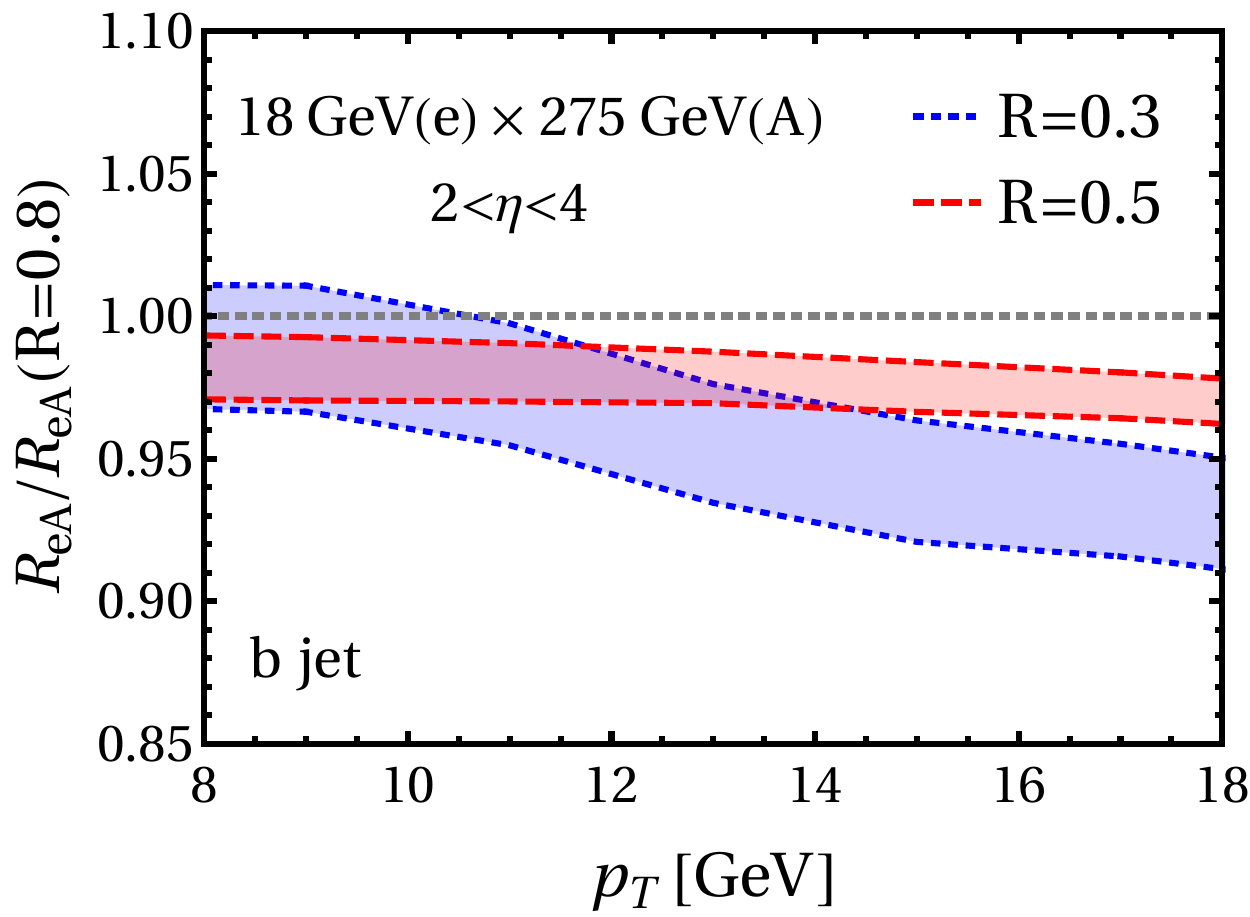} 
	\caption{ The ratio of $R_{eA}$ normalized by $R_{eA}(R=0.8)$ for c-jets and b-jets as a function of $p_T$ at the EIC. Blue bands (dotted lines) and red bands (dashed lines) correspond to  $R=0.3$ and $R=0.5$, respectively. The kinematic regions and beam energy combinations are as in Fig.~(\ref{fig:bcj_EIC}). }
	\label{fig:bcj_EIC_nPDF}
\end{figure}

\begin{figure}
	\centering
	\includegraphics[width=0.39\textwidth]{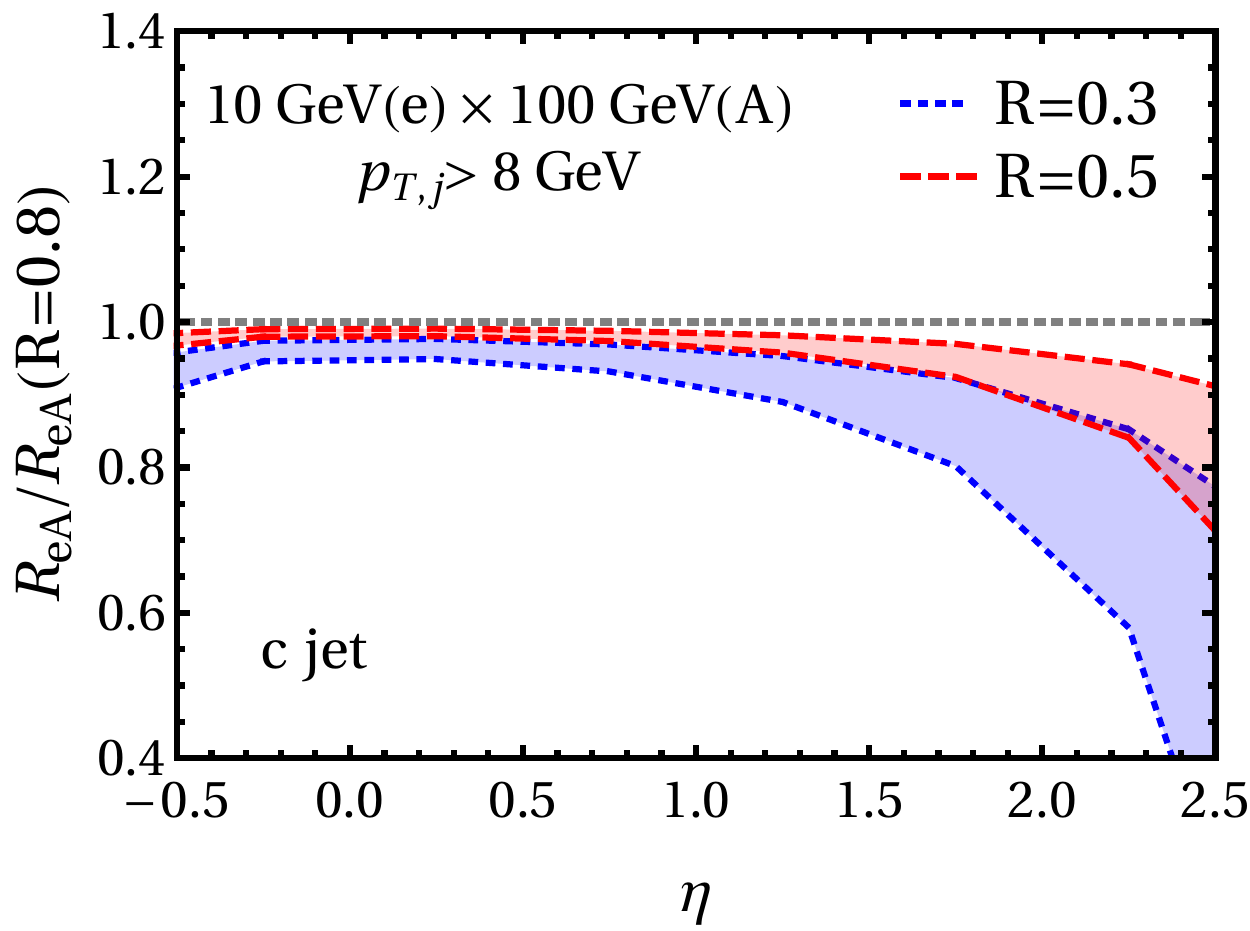}\,\,\,
	\includegraphics[width=0.39\textwidth]{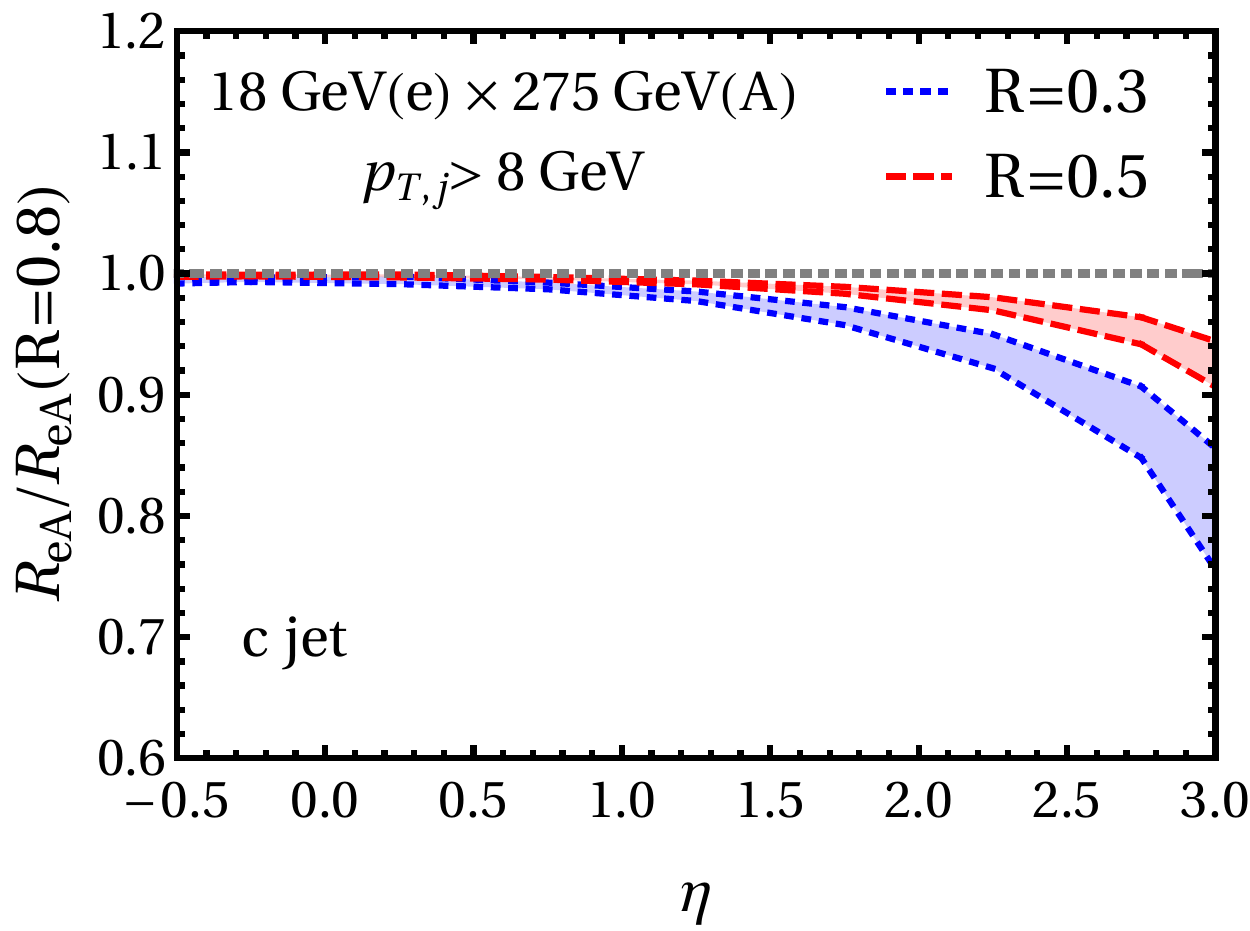} \\	
	\includegraphics[width=0.39\textwidth]{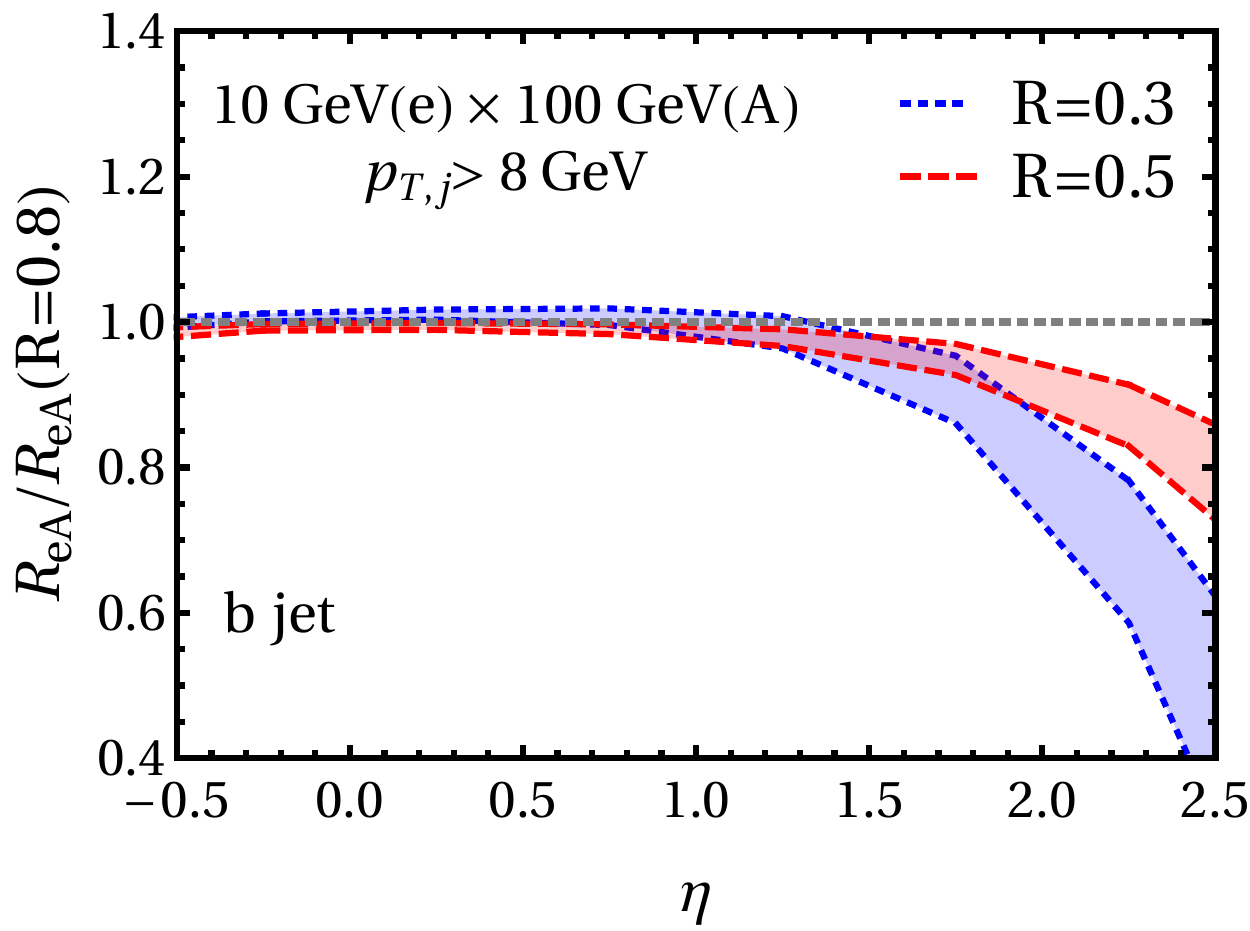}\,\,\,
	\includegraphics[width=0.39\textwidth]{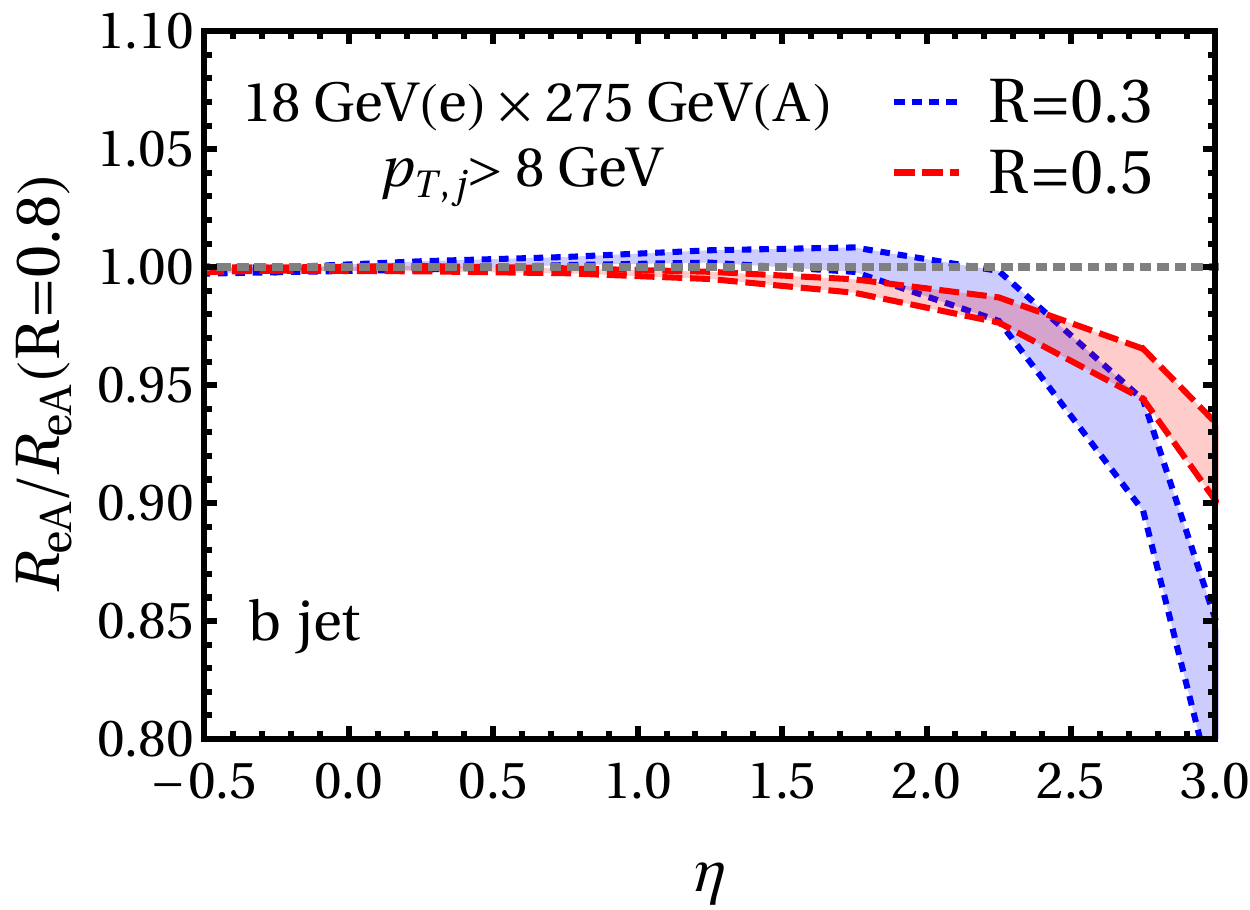} 
	\caption{ The ratio of $R_{eA}$ normalized by $R_{eA}(R=0.8)$ for heavy flavor jet production as a function of $\eta$ at the EIC. Blue bands (solid lines) and red bands (dashed lines) correspond to  $R=0.3$ and $R=0.5$, respectively. }
	\label{fig:hj_eta_nPDF}
\end{figure}

In the following numerical calculations we use CT10nlo PDF sets~\cite{Lai:2010vv}  for the proton and the nCTEQ15 PDF sets~\cite{Kovarik:2015cma} for the nucleus, and the associated strong coupling provided by  {\sc  Lhapdf6}~\cite{Buckley:2014ana}.  The medium-induced splitting kernel up to first order of opacity~\cite{Ovanesyan:2011kn,Kang:2016ofv,Sievert:2019cwq} were used, consistent with earlier EIC studies~\cite{Li:2020rqj,Li:2020zbk}.  
As a default choice, the nominal transport coefficient of cold nuclear matter for quarks is set to be $\langle q_\perp^2 \rangle /\lambda_q = 0.05$ GeV$^2$/fm from the
above references. The theoretical uncertainties in this section are evaluated by varying the transport parameter up and down by a factor of two, which represents the sensitivity of the observable to the transport coefficient.

To investigate the nuclear medium effects,  we study the ratio of the cross sections in electron-gold (e+Au) collisions normalized by the number of nucleons to the one in e+p collisions. 
\begin{equation}
R_{eA}(R)=\frac{1}{A} {\int \frac{d\sigma}{d \eta d p_T}   \big |_{e+A} \, d\eta } \; \Bigg / \; {\int \frac{d\sigma}{d \eta d p_T} 
 \big |_{e+p} \, d\eta   } \, .
\end{equation}
In Figs.~\ref{fig:bcj_EIC} we present $R_{eA}$ for c-jets and b-jets as a function of the transverse momentum $p_T$ in the laboratory frame in two rapidity bins 0$<\eta<$2 and 2$<\eta<$4.   The left column of panels is for 10 GeV $\times$ 100 GeV e+Au collision and the right column of panels  is for 18 GeV $\times$ 275 GeV ones. 
 The in-medium shower corrections induced by the  interactions between the final-state partons and the nucleus vary with the parton energy in the nuclear rest frame, where the lower energy parton receives larger medium corrections. Therefore, in the  forward (nucleus-going) rapidity region  $2<\eta<4$ we can see more significant jet quenching due to final-state interactions in the large nucleus. Furthermore, a clear separation of jet suppression is observed as a function of the radius $R$.  Initial-state effects reflecting the difference between proton and nuclear PDFs also play an important role in $R_{eA}$. In the kinematic domains that we consider the smaller beam energies combination is primarily sensitive to the so called EMC region. At the higher beam energies combination we see a clear transition from the anti-shadowing region near midrapidity to the EMC at forward rapidity. Initial-state effects are large for c-jets and b-jets as their production channels are dominated by gluon and sea quarks.

To understand the structure and evolution of showers containing heavy quarks in cold nuclear matter and to use them as tomographic probes at the EIC, it is essential to reduce the effects of nPDFs and enhance the effects of final-state interactions. A successful strategy was developed on the example of inclusive light parton jets~\cite{Li:2020zbk} and it involves measuring the ratio of the modifications with different jet radii, e.g. $R_{eA}(R)/R_{eA}(R = 0.8)$.  In such double ratio
initial-state effects in e+A reactions will cancel for jets with a similar kinematics. This double ratio is also an observable sensitive to the angular distribution of in-medium branching processes~\cite{Vitev:2008rz,Li:2020rqj}. Furthermore, it provides an opportunity to explore smaller center-of-mass energies where the final-state effects are expected to be sizable even though the cross section is small. Such measurements will take advantage of the high-luminosity design of the future facility. Our predictions for the double ratio of jet cross section suppression in two rapidity bins at the EIC are presented in Fig.~\ref{fig:bcj_EIC_nPDF}, where the left and right panels correspond to the results for 10 GeV (e) × 100~GeV~(A) and 18 GeV (e) × 275 GeV (A) collisions. The blue and red bands correspond to jet radii $R=$0.3 and 0.5, respectively, with the large normalization radius $R=0.8$ and with variation of the cold nuclear matter transport coefficient. Since medium-induced parton showers are broader than the ones in the vacuum, for smaller jet radii the suppression from final-state interactions is more significant. For both c-jets and b-jets, we can  identify  larger in-medium effects at 10 GeV $\times$ 100 GeV e+Au collision than at 18 GeV$\times$ 275 GeV collision  with the same $p_T$ range fixed. In fact, these are as large as the ones observed for light jets. Additionally, jet production in the forward rapidity region $2<\eta<4$ receives the largest in-medium corrections. 

We flesh out the  rapidity dependence of $R_{eA}$ in Fig.~\ref{fig:hj_eta_nPDF}. Instead of rapidity,  we have integrated the  $p_T$-dependent cross sections above 8~GeV. 
Once again  we show 10 GeV$\times$ 100 GeV (left) and 18 GeV$\times$ 275 GeV e+Au collisions (right). The upper and bottom two panels  correspond to c-jets and b-jets, respectively. 
It is very clear that the medium-induced suppression is much enhanced in the forward rapidity region where the
jet has smaller energy in the nuclear rest frame - a region that should
be well-instrumented for these key measurements at the EIC.

\subsection{Heavy flavor-tagged jet substructure}
\begin{figure}[t]
	\centering
	\includegraphics[width=0.42\textwidth]{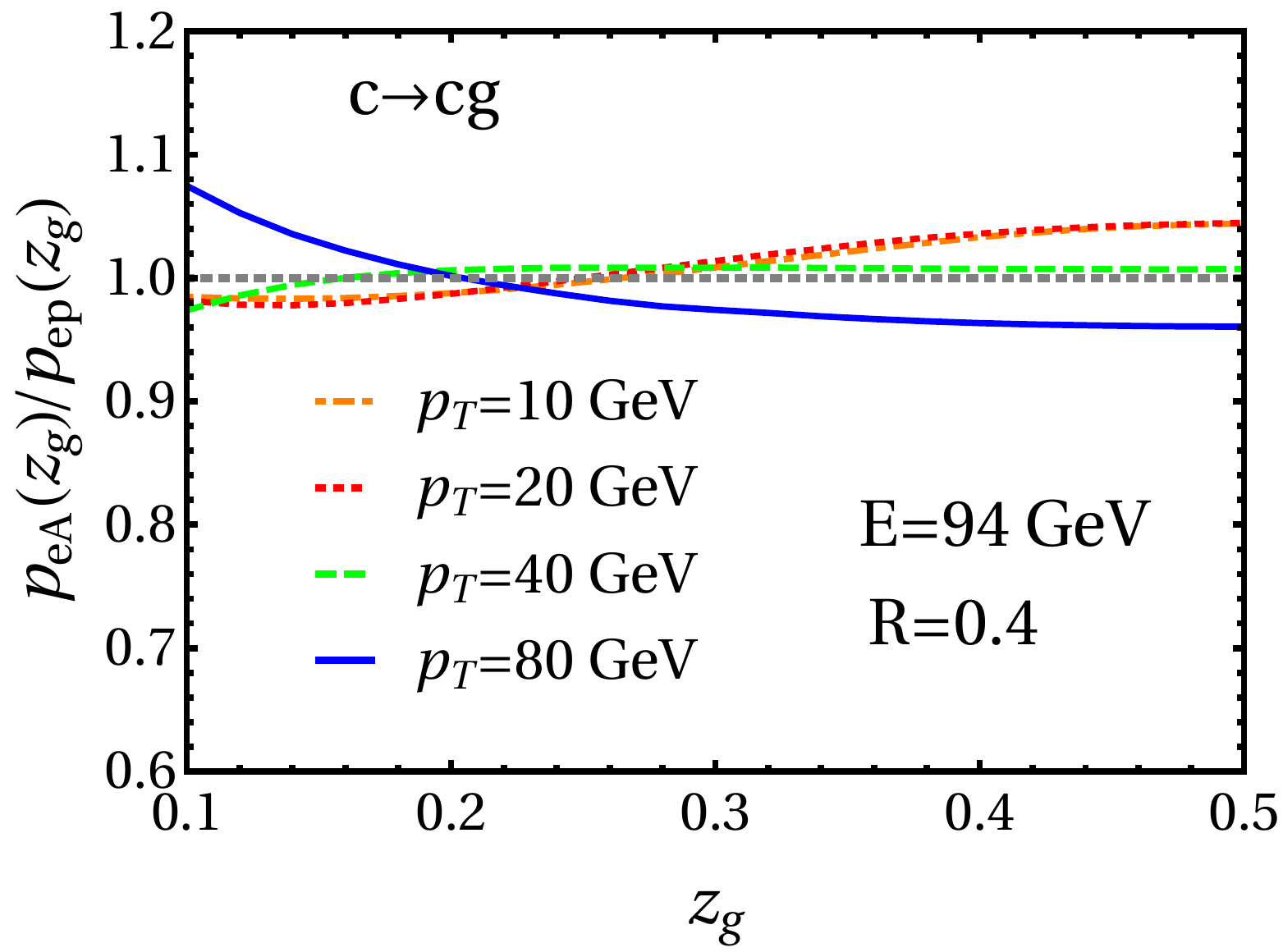}\,\,\,
	\includegraphics[width=0.42\textwidth]{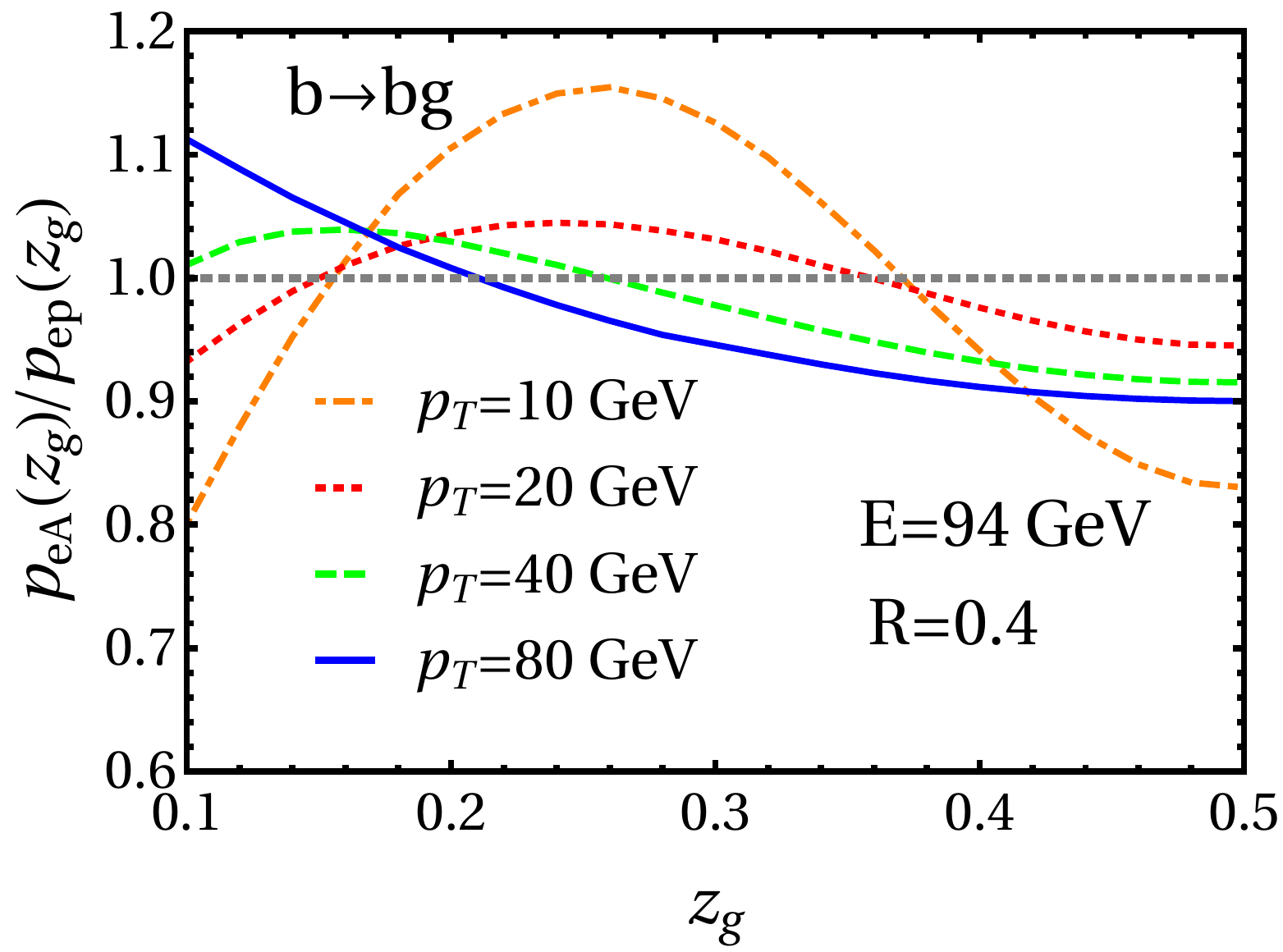} 
	\caption{The modification of the jet splitting functions for $c\to cg$ (left) and $b\to bg$ (right). The energy of the jet is fixed at 94 GeV at the rest frame of the nucleus. The orange dot-dashed, red dotted, green dashed and blue solid lines denote the distribution with $p_T$= 10 GeV, 20 GeV, 40 GeV and 80 GeV, respectively. }
	\label{fig:zg_pt}
\end{figure}

The jet splitting-function observable in e+A, i.e. the distribution of $z_{g}$, only depends on the final-state interactions between the jet and cold nuclear matter. At the EIC we probe a regime very different from heavy ion collisions. To illustrate this we study the $p_{T}$ dependence of $p_{eA}(z_g)/p_{ep}(z_g)$ and fix the jet energy in the rest frame of the nucleus to be 94 GeV for a jet radius $R=0.4$.  With the design energies, the transverse momentum of jets at EIC for practical purposes will be limited to about 30 GeV. However, a much larger range of jet transverse momentum $10<p_T<80$ GeV  is used here to demonstrate the $p_T$ dependence and compare the large $p_T$ region with the previous analyses in Ref.~\cite{Li:2017wwc}.\footnote{It
is completely clear that the higher transverse momenta shown cannot be
reached at EIC kinematics. That would be possible in heavy ion collisions.
We ignore kinematic constraints to make a physics point.} The results
for charm-quark jets (left) and bottom-quark jets (right) are presented in
Fig.~\ref{fig:zg_pt} for the nominal value of the medium's transport parameter.
Even though at large jet $p_{T}$ the modifications of c-jet and b-jet substructure
are similar because $m \ll p_{T}$, we are beginning to see hints of the
quark mass effect. When $p_{T}\sim E$, the results have a qualitatively
similar behavior with the $z_{g}$ modifications in heavy-ion collisions~\cite{Li:2017wwc}.
 For the small jet $p_T$ a much larger mass effect is observed  by comparing the c-jet and b-jet modifications.  Not only is the magnitude larger for the bottom quark-initiated  ones, but the shape of the modification differs  consistent with the fact that the scale $z_g m$ plays an important role in the branching Eq.~(\ref{eq:Msp1}). 

\begin{figure}[ht]
	\centering
	\includegraphics[width=0.42\textwidth]{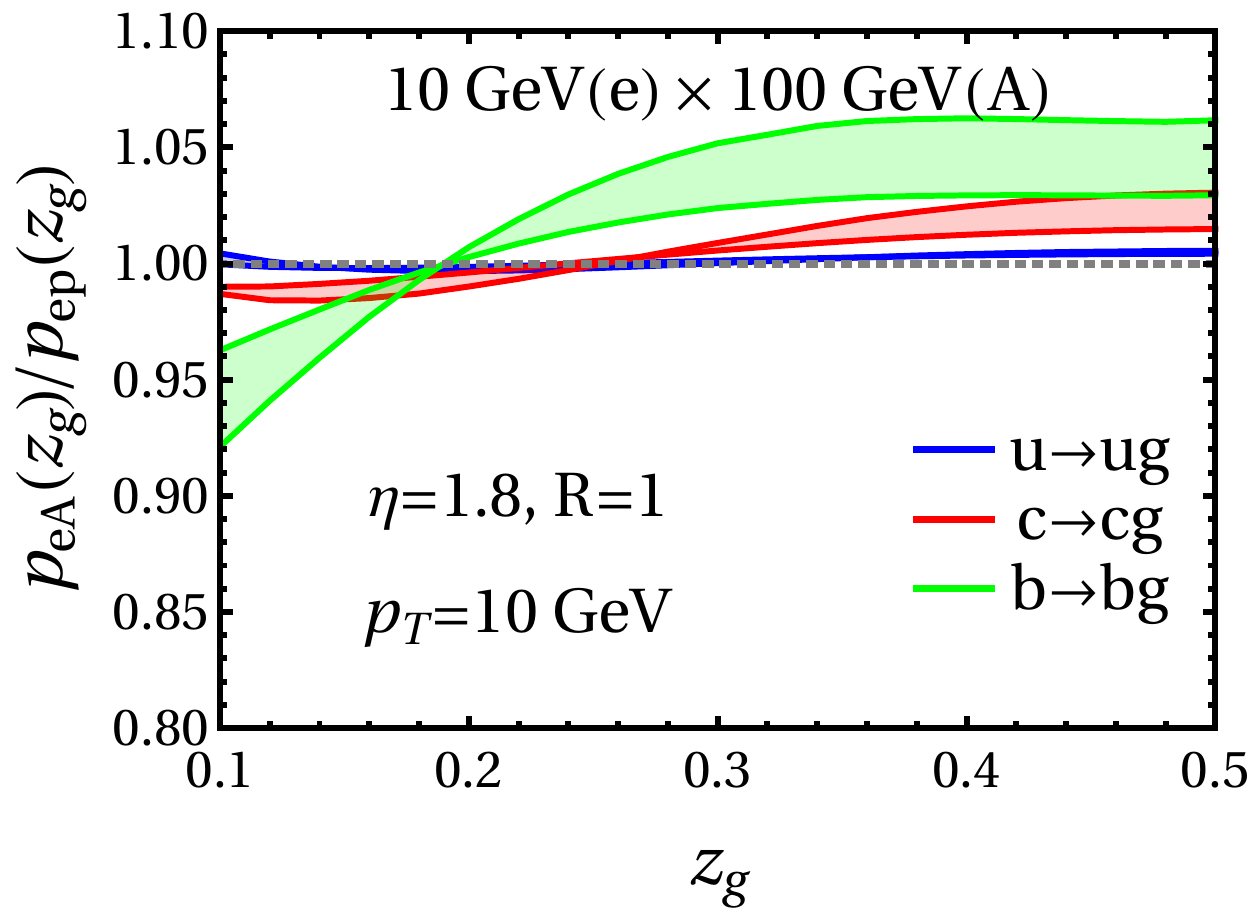}\,\,\,
	\includegraphics[width=0.42\textwidth]{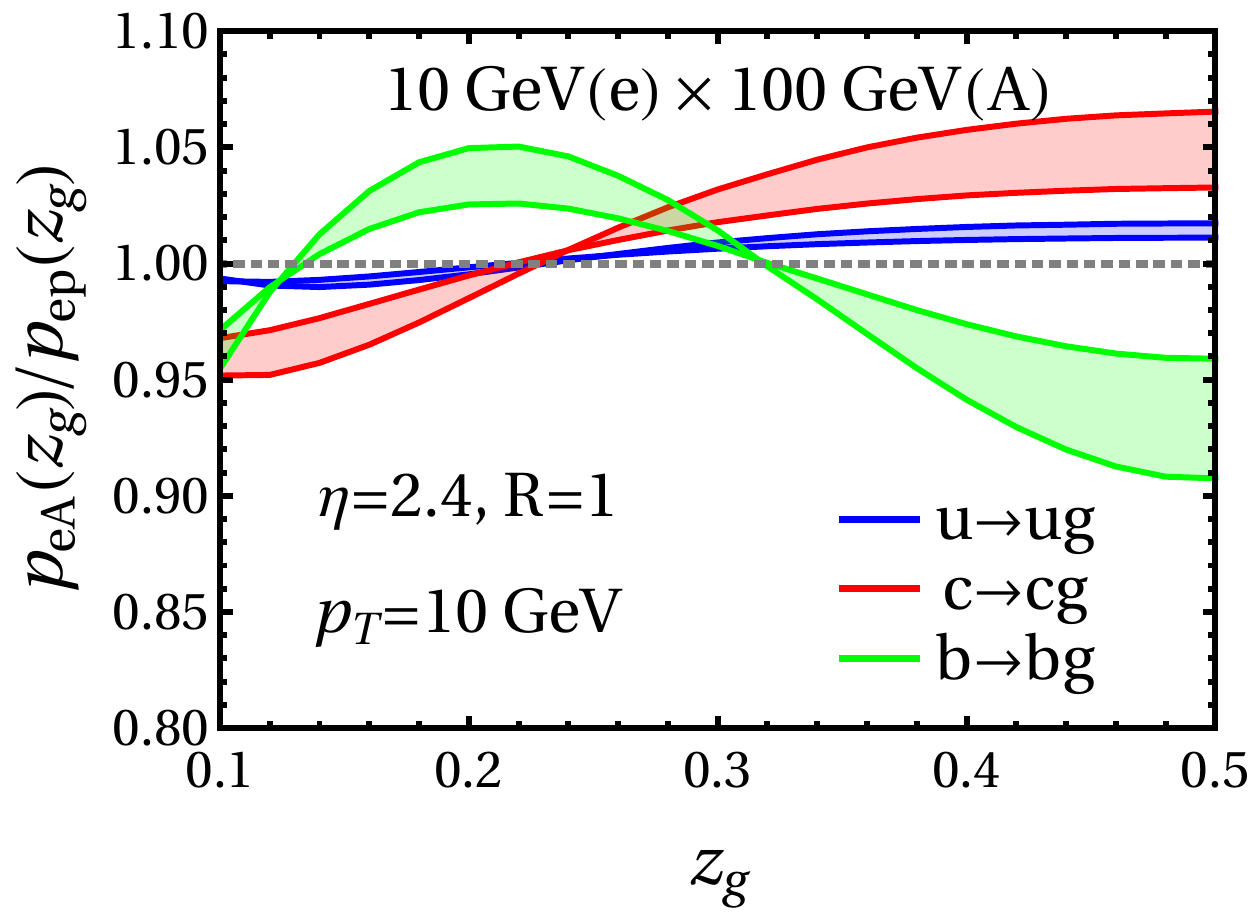} \\	
	\includegraphics[width=0.42\textwidth]{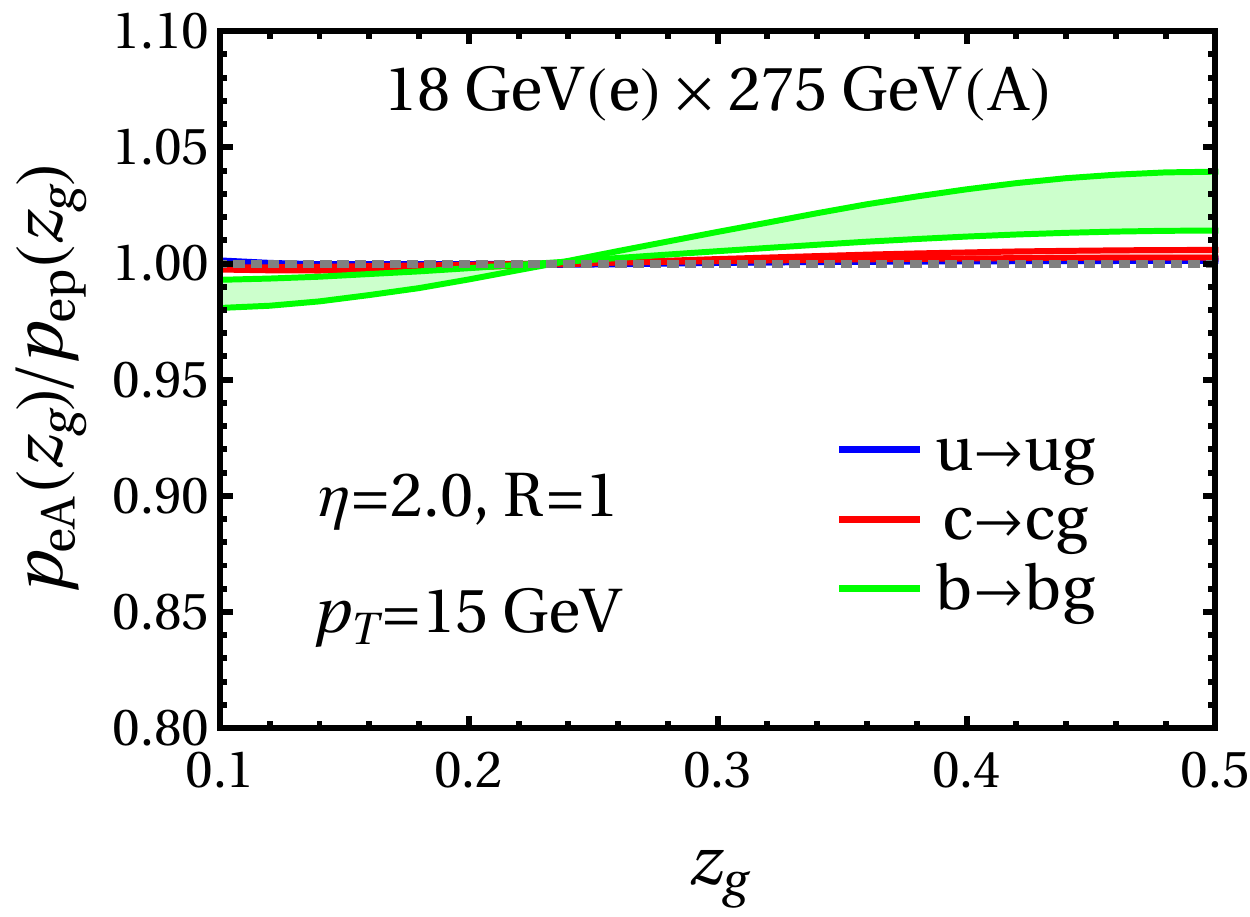}\,\,\,
	\includegraphics[width=0.42\textwidth]{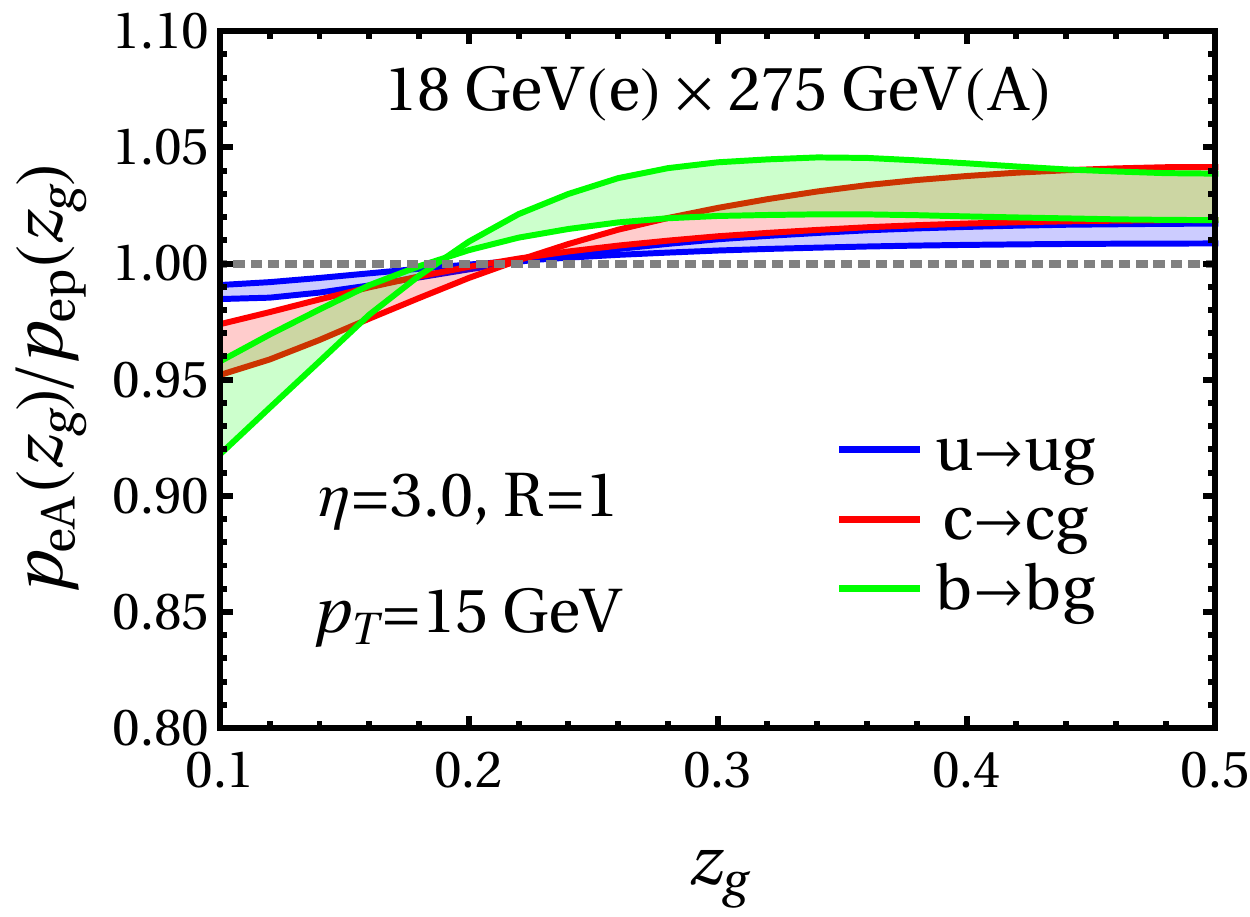} 
	\caption{The modification of the jet splitting functions for c-jets and b-jets vs $z_g$ at the EIC. The upper and bottom two panels correspond to the result for 10 GeV $\times$ 100 GeV and  18 GeV $\times$ 275 GeV e+Au to e+p  collisions, respectively. The jet radius $R=1$ and two different rapidities are presented for each energy combination.}
	\label{fig:zg_eta}
\end{figure}

In Fig.~\ref{fig:zg_eta}, we present the modification of the jet splitting functions for charm jets and bottom jets with fixed jet rapidity at the EIC. Upper panels correspond to the results for 10 GeV $\times$ 100 GeV collisions  and bottom  panels correspond to the results  18 GeV $\times$ 275 GeV collisions, respectively. The jet radius is chosen to be $R=1$,  thus there is still available, albeit limited, phase space to ensure $m\ll p_T R$. The blue, red and green bands correspond to our calculations for light, c-quark and b-quark jets, respectively, and the bands reflect the uncertainties to the variation of the transport parameter by a factor of two. It is clear that, just as in the case of the cross sections, the in-medium corrections  are larger in the forward rapidity region of the jet, because the corresponding jet energy in the rest frame of the nucleus is smaller. At the transverse momenta accessible at the EIC the modification is quite different when compared to heavy ion collisions. 
By changing rapidity we can also see a difference
in the modification pattern of bottom-quark jets which we attribute to
the interplay between mass and parton energy in the non-Abelian Landau-Pomeranchuk-Migdal
(LPM) effect for heavy quarks. These findings are intriguing and, clearly,
more detailed future studies of heavy flavor-tagged jet substructure at
the EIC will be quite important.

\section{Conclusions}\label{conclusions}

In summary, we presented the first calculation of semi-inclusive charm-quark jet and bottom-quark jet production and substructure in e+A relative to e+p collisions at the EIC. Our formalism allowed to obtain NLO results by consistently combining the parton level cross sections and semi-inclusive jet functions up to NLO, and included resummation for small jet radii in electron-hadron reactions. We found  that heavy flavor-tagged jet production is more sensitive to the gluon and sea quark distributions in nucleons and nuclei in comparison to light jets. Thus, in kinematic regions where $R_{eA}$ is dominated by initial-state  nPDF effects the modification was even stronger when compared to inclusive jets. Similar to the case of light jets,  by applying the strategy of studying ratios of the nuclear modification with two different jet radii we were successful in eliminating nPDF effects, primarily the anti-shadowing and the EMC effect in the regions of interest.  The remaining quenching of the jet spectra can be as large as a factor of two for small jet radii, for example $R=0.3$, and can clearly be attributed to final-state interactions and in-medium modification of parton showers containing heavy quarks. This suppression is  comparable to the one predicted for light jets and expected to be observed in the proton/nucleus going direction. In contrast, near mid rapidity and at backward rapidity the  deviation of $R_{eA}(R)/R_{eA}(R=0.8)$  from unity is small since the energy of the parton/jet in the rest frame of the nucleus is very large. This, in turn, strongly reduces the contribution of in-medium parton shower due to the non-Abelian LPM effect. In fact, even at forward rapidity and smaller center-of-mass energies  the parton energies in nuclear rest frame are quite sizeable and, therefore, there isn't much difference in the suppression of c-jets and b-jets.

We complemented the calculation of semi-inclusive jet cross sections with a calculation of the groomed, soft-dropped momentum sharing distribution. Our results show that the substructure modification in e+A relative to e+p reactions is relatively small -- on the order of 10\% or smaller. Still, just like in the case of heavy ion collisions at relatively small transverse momenta the differences in the subjet distribution are most pronounced for b-jets, followed by c-jets. In the kinematic regime accessible at the EIC the modification of light jets was found to be the smallest.  In contrast to the heavy ion case, however, there is significant difference between the energy of the parton in the rest frame of the nucleus  and the jet scale which determines the available phase space for substructure even for large radii $R\sim 1$. Thus, the  jet  momentum sharing distribution at the  EIC probes a different interplay between the heavy quark mass and  suppression of small-angle medium-induced radiation  -- a regime that can only be accessed at the EIC and merits further investigation in the future. We conclude by pointing out that with the theoretical tools that are becoming available one can also look at how subeikonal corrections to in-medium branching, such as the effects of varying matter density~\cite{Sadofyev:2021ohn}, propagate into the observables that we predicted in this work. 

\section*{Acknowledgments}
This work was supported by the U.S. Department of Energy under Contract No. 89233218CNA000001, the Los Alamos National Laboratory LDRD program, and the TMD topical collaboration for nuclear theory. H.T. Li is supported  by   the U.S. Department of Energy under Contract  No. DE-AC02-06CH11357 and the National Science Foundation under Grant No. NSF-1740142.

\bibliographystyle{JHEP}
\bibliography{mybib}

\end{document}